\documentclass[oneside,10pt]{amsart}
\pdfoutput=1

\usepackage{geometry}
\usepackage[T1]{fontenc}
\usepackage[utf8]{inputenc}
\usepackage{lmodern}
\usepackage[dvipsnames]{xcolor}
\usepackage{tikz,tikz-3dplot}
\usetikzlibrary{positioning,arrows.meta,decorations.pathreplacing}

\makeatletter
 \tikzoption{canvas is plane}[]{\@setOxy#1}
 \def\@setOxy O(#1,#2,#3)x(#4,#5,#6)y(#7,#8,#9)%
   {\def\tikz@plane@origin{\pgfpointxyz{#1}{#2}{#3}}%
    \def\tikz@plane@x{\pgfpointxyz{#4}{#5}{#6}}%
    \def\tikz@plane@y{\pgfpointxyz{#7}{#8}{#9}}%
    \tikz@canvas@is@plane
   }
\makeatother  

\newsavebox{\smallclawbox}
\savebox{\smallclawbox}{%
\begin{tikzpicture}[x=1ex,y=1ex,baseline={([yshift=-.7ex]current bounding box.center)}]
    \coordinate (v) at (0,0);
    \coordinate (v1) at ([shift=(90:.8)]v);
    \coordinate (v2) at ([shift=(210:.8)]v);
    \coordinate (v3) at ([shift=(330:.8)]v);
    \draw (v) -- (v1);
    \draw (v) -- (v2);
    \draw (v) -- (v3);
    \filldraw (v) circle (.6pt);
    \filldraw (v1) circle (.6pt);
    \filldraw (v2) circle (.6pt);
    \filldraw (v3) circle (.6pt);
\end{tikzpicture}%
}
\newcommand{\smallclaw}{{\usebox{\smallclawbox}}}

\newcommand{\CC}{{\mathbb C}}

\newcommand{\FF}{{\mathbb F}}

\newcommand{\QQ}{{\mathbb Q}}
\newcommand{\RR}{{\mathbb R}}

\newcommand{\ZZ}{{\mathbb Z}}
\newcommand{\ee}{\mathrm{e}}
\newcommand{\ii}{\mathrm{i}}

\newcommand{\dd}{\mathrm{d}}

\newcommand{\mm}{{\overline{m}}}
\newcommand{\zz}{{\overline{z}}}
\newcommand{\Gg}{{\overline{G}}}

\newcommand{\sE}{\mathcal{E}}

\newcommand{\sL}{\mathcal{L}}

\newcommand{\sO}{\mathcal{O}}
\newcommand{\sV}{\mathcal{V}}

\renewcommand{\Re}{\mathrm{Re}\,}
\renewcommand{\Im}{\mathrm{Im}\,}

\newcommand{\intsv}{\int_\mathrm{sv}}

\theoremstyle{plain}

\title{Seven loops $\phi^4$}
\author{Oliver Schnetz}
\address{Department Mathematik\\
Cauerstra{\ss}e 11\\
91058 Erlangen, Germany}
\email{schnetz@mi.uni-erlangen.de}

\begin{document}
\begin{abstract}
We give a detailed account of the theory of position space renormalization using graphical functions in the case of dimensionally regularized $\phi^4$ theory in four dimensions.
In this theory we calculate the beta function, the mass gamma function, and the self energy to seven loops in the minimal subtraction scheme.
The anomalous dimension $\gamma$ is calculated to loop order eight.

When possible, we generalize to even dimensions $\geq4$ with particular focus on $\phi^3$ theory in six dimensions. In this theory we calculate the anomalous dimension $\gamma$ to
loop order six.
\end{abstract}
\maketitle

\section{Introduction}
The calculation of the renormalization functions $\beta$, $\gamma$, and $\gamma_m$ for a given Quantum Field Theory (QFT) is a classical problem in particle physics \cite{IZ}.
The beta function determines the running of the coupling, whereas $\gamma$ and $\gamma_m$ are the anomalous dimensions of the field and the mass.
Knowledge of the renormalization functions also provides approximations for the critical exponents of phase transitions in certain universality classes of statistical models \cite{ZJ}.
So, the calculation of $\beta$, $\gamma$, and $\gamma_m$ to highest possible loop orders has impact beyond QFT.

The classical method to calculate these renormalization functions is based on momentum space. In many decades of research very refined techniques were developed, most notably
the reduction to master integrals after infrared $R^\ast$ reduction. An overview over the classical methods is in \cite{KP6loopbeta} where the beta function in $\phi^4$ theory
is calculated to six loops (i.e.\ six independent cycles of the underlying Feynman graphs). In $\phi^4$ theory a spin zero boson interacts with itself in a quartic process
(in Nature realized in the Higgs sector of the standard model). The six loop result is world leading within all QFTs for the classical method. Outside $\phi^4$ theory, the beta function
is typically known to five loops \cite{5lphi3,classphi3,5lQCD1,5lQCD2,5lQCD3}.

A decade ago the theory of graphical functions was developed by the author to calculate Feynman periods in four dimensions (residues of Feynman integrals with
a single logarithmic divergence) \cite{gf}. Using the theory of graphical functions it was possible for the first time to calculate a significant number of Feynman periods.
With these results the author discovered a connection between QFT and motivic Galois theory, a deep and rich mathematical structure which generalizes Galois theory to higher dimensions
in the context of algebraic integrals (see e.g.\ \cite{Bcoact1,Bcoact2}). The discovery lead to the formulation of the coaction conjectures in \cite{coaction}. Now, the connection (also named `the
coaction principle' or `the cosmic Galois group') has turned into a classic field in modern QFT (see e.g.\ \cite{Cosmic} and the references therein).

In collaboration with Michael Borinsky, it has lately been possible to generalize the theory of graphical functions to even dimensions $\geq4$ \cite{gfe} (the article also serves
as an up to date reference for graphical functions in four dimensions). While for $\phi^4$ theory we only need results in four dimensions, basically any other QFT (at least conveniently) uses
higher even dimensions: The $\phi^3$ theory of spin zero bosons with a cubic self-interaction lives in six dimensions and calculations in gauge theories map by a dimensional shift mechanism
to graphical functions in four, six, and eight dimensions. In $\phi^3$ theory, the graphical function method provides a complete list of Feynman periods up to six loops and partial results
up to nine loops \cite{phi3,Shlog}. All results are consistent with the connection to motivic Galois theory.

Every graphical function is a single-valued function on the complex plane with singularities at 0 and 1 (and at infinity) \cite{par}.
Calculations within the theory of graphical functions showed demand to understand a new class of hyperlogarithmic functions, the generalized single-valued hyperlogarithms (GSVHs) \cite{GSVH}.
In contrast to standard hyperlogarithms where singularities correspond to punctures in the complex plane, GSVHs allow the integrand to have a more complicated structure. In the denominator
of the integrand factors are possible which are bilinear in the integration variable and its complex conjugate. These factors $az\zz+bz+c\zz+d$, $a,b,c,d\in\CC$, generalize
the factors $z-e$ or $\zz-f$, $e,f\in\CC$, in standard single-valued hyperlogarithms \cite{BrSVMP}. For single-valuedness one needs the condition that before integration the non-point-like zeros
in the denominator of a GSVH has to be lifted by zeros in the numerator (although the factor does not cancel in the fraction, see \cite{GSVH} for a precise statement).
This condition is a general property of graphical functions. The most prominent example of a non-standard factor in the denominator is $z-\zz$. The denominator $z-\zz$ was already studied
in \cite{Duhr} with no direct connection to graphical functions.

After the establishment of the theory of graphical functions in \cite{gfe} and the theory of GSVHs in \cite{GSVH}, the final step to more physical applications is the generalization
to non-integer dimensions which is required by the commonly used dimensional regularization scheme. With this extension to $d-\epsilon$ dimensions it becomes possible to handle Feynman integrals
with subdivergences.

The general idea is reasonably straight forward. Graphical functions in $d-\epsilon$ dimensions have Laurent expansions in $\epsilon$ where each coefficient resembles the properties
of graphical functions in integer dimensions.
In particular, each coefficient is a single-valued function on $\CC\backslash\{0,1\}$. At reasonably low loop orders the function space of the coefficients is covered by GSVHs with
ubiquitous use of the denominator $z-\zz$. The main technique of appending an edge to a known graphical function generalizes to $d-\epsilon$ dimensions (Section 4).

Still, there are many subtleties which have to be worked out. On one hand one has to handle singularities, which is very efficiently facilitated
by a subtraction scheme (see Section \ref{sectreg}). Only with this regularization prescription one is able to transfer the basic tools for graphical functions to non-integer dimensions
(see Sections \ref{sectcon} and \ref{sectid}). On the other hand there exists a plethora of new techniques which are needed to make efficient use of graphical functions
in non-integer dimensions (see Sections \ref{sectapp} and \ref{sectrerou}). Small graphs whose graphical functions are not accessible by any known method can sometimes be
evaluated with E. Panzer's HyperInt which was suggested by F. Brown \cite{HyperInt} (see Section \ref{secthyp}).
We begin the article with a section on graphical functions (Section \ref{sectgraph}) and on their use for the calculation of renormalization functions (Section \ref{sectren}).
In Section 6 we explain how to calculate Feynman periods from graphical functions in $d-\epsilon$ dimensions.
We end the article with a worked example that demonstrates how the various techniques interplay to calculate the Laurent coefficients of a rather complex graphical function in four dimensions
(Section \ref{sectex}).

All three theories -- graphical functions, GSVHs, and the extension of graphical functions to non-integer dimensions -- are new and very different from momentum space techniques.
In particular, the presented techniques for the calculation of graphical functions in non-integer dimensions are far from exhaustive. Moreover, the implementation {\tt HyperlogProcedures}
\cite{Shlog} is written in Maple with no good handling for systems of linear equations relating coefficients of graphical functions for various graphs and various powers of $\epsilon$.
This limits the power of the new technique (see e.g.\ \cite{phi3} for a C$++$ addition). It is to be expected that the method will become much
more powerful with more systematic approaches in the future. Still, the theory proved immediately successful. It quickly was possible to confirm
the six loop momentum space result by M.V. Kompaniets and E. Panzer \cite{KP6loopbeta}. Soon thereafter, a new result for the seventh loop order of the $\phi^4$ beta function
was available \cite{numfunct}. In the meantime, the gamma function in $\phi^4$ theory has been calculated to eight loops in the minimal subtraction scheme,
\begin{eqnarray}\label{8loopgamma}
\gamma_8(\phi^4)&=&\frac{1506066907}{6635520}+\frac{115767719}{414720}\zeta(3)+\frac{549949}{829440}\pi^4-\frac{1274869}{11520}\zeta(5)+\frac{39437}{90720}\pi^6
+\frac{642917}{1920}\zeta(3)^2\nonumber\\
&&+\,\frac{1837}{8640}\pi^4\zeta(3)-\frac{6018361}{5760}\zeta(7)+\frac{672397}{6480000}\pi^8+\frac{807}{2}\zeta(3)\zeta(5)+\frac{3801}{100}\zeta(5,3)+\frac{635}{2268}\pi^6\zeta(3)\nonumber\\
&&-\,\frac{523}{360}\pi^4\zeta(5)-\frac{169}{2}\zeta(3)^3-\frac{608849}{432}\zeta(9)+\frac{22553}{1964655}\pi^{10}+\frac{8}{15}\pi^4\zeta(3)^2-\frac{10429}{21}\zeta(5)^2\nonumber\\
&&+\,\frac{105}{2}\zeta(3)\zeta(7)-\frac{94}{7}\zeta(7,3)\\
&\approx&1171.87570158\ldots,\nonumber
\end{eqnarray}
where
$$
\zeta(n)=\sum_{k\geq1}\frac{1}{k^n},\qquad\zeta(n_1,n_2)=\sum_{k_1>k_2\geq1}\frac{1}{k_1^{n_1}k_2^{n_2}}.
$$
In $\phi^3$ theory the five loop result in \cite{phi3} was obtained using graphical functions. Now, the calculation of the six loop result for the gamma function has been finalized,
\begin{eqnarray}\label{6loopgamma}
\gamma_6(\phi^3)&=&\frac{29506113557}{9674588160}+\frac{574643}{46656}\zeta(3)+\frac{1378253}{223948800}\pi^4+\frac{708913}{77760}\zeta(5)+\frac{25637}{7838208}\pi^6
-\frac{11333}{10368}\zeta(3)^2\nonumber\\
&&-\,\frac{13}{14400}\pi^4\zeta(3)-\frac{25967}{2304}\zeta(7)+\frac{209}{324000}\pi^8-\frac{2}{3}\zeta(3)\zeta(5)-\frac{21}{10}\zeta(5,3)-\zeta(3)^3-\frac{1567}{72}\zeta(9)\\
&\approx&-0.331182447708\ldots.\nonumber
\end{eqnarray}
Note that, somewhat unexpectedly, $\gamma_6(\phi^3)$ is negative.

The author currently works on a six loop result for the $\phi^3$ beta function. A full eight loop result for the $\phi^4$ beta functions is still out of reach.

The method of graphical functions can also be used to calculate renormalization functions of gauge theories. However, the application to gauge theories has not yet been
implemented. It is expected that some work is necessary to fully utilize the theory for the calculation of renormalization functions in QED or QCD. We expect
that six loop results should be possible in the future.

\section*{Acknowlegements}
The author is grateful to E. Panzer and M. Borinsky for many very valuable discussions. During most of the work on this project the author was supported by DFG grant SCHN~1240.

\section{Graphical functions}\label{sectgraph}
\begin{figure}
\tdplotsetmaincoords{80}{120}
\centering
\begin{tikzpicture}[tdplot_main_coords,scale=.85]
\draw[thin, ->,black!50] (0,0,0) -- (4.5,0,0) node[anchor=south east,opacity=1]{$x^1$};

\draw[thin, ->,black!50] (0,0,0) -- (0,4.5,0)  node[anchor=south west,opacity=1]{$x^2$};

\draw[thin, ->,black!50] (0,0,0) -- (0,0,4.5)  node[anchor=south west,opacity=1]{$x^D$};

\tdplotsetrotatedcoords{0}{90}{0};
\draw[dotted,black!50,tdplot_rotated_coords] (0,.8,0) arc (90:180:.8);

\pgfmathsetmacro{\Ox}{32}
\pgfmathsetmacro{\Oy}{14}
\pgfmathsetmacro{\Oz}{3}
\pgfmathsetmacro{\Onex}{-5}
\pgfmathsetmacro{\Oney}{3}
\pgfmathsetmacro{\Onez}{1}
\pgfmathsetmacro{\Zx}{-12}
\pgfmathsetmacro{\Zy}{1}
\pgfmathsetmacro{\Zz}{6}

\tdplotcrossprod(\Onex,\Oney,\Onez)(\Zx,\Zy,\Zz)
\pgfmathsetmacro{\rx}{\tdplotresx}
\pgfmathsetmacro{\ry}{\tdplotresy}
\pgfmathsetmacro{\rz}{\tdplotresz}
\tdplotcrossprod(\rx,\ry,\rz)(\Onex,\Oney,\Onez)
\pgfmathsetmacro{\tx}{\tdplotresx/100}
\pgfmathsetmacro{\ty}{\tdplotresy/100}
\pgfmathsetmacro{\tz}{\tdplotresz/100}

\pgfmathsetmacro{\dplane}{\rx*\Ox + \ry*\Oy + \rz*\Oz}

\pgfmathsetmacro{\OneLen}{sqrt(\Onex*\Onex+\Oney*\Oney+\Onez*\Onez)};
\pgfmathsetmacro{\iLen}{sqrt(\tx*\tx+\ty*\ty+\tz*\tz)};

\tikzset{perspective/.style= {canvas is plane={O(0,0,0)x(\Onex/\OneLen,\Oney/\OneLen,\Onez/\OneLen)y(\tx/\iLen,\ty/\iLen,\tz/\iLen)}} }
\pgfmathsetmacro{\axisscale}{2};
\tikzset{perspective2/.style= {canvas is plane={O(0,0,0)x(\axisscale*\Onex/\OneLen,\axisscale*\Oney/\OneLen,\axisscale*\Onez/\OneLen)y(\axisscale*\tx/\iLen,\axisscale*\ty/\iLen,\axisscale*\tz/\iLen)}} }

\pgfmathsetmacro{\axisovershoot}{2};
\pgfmathsetmacro{\axisundershoot}{.5};

\coordinate (v0) at (\Ox,\Oy,\Oz);
\coordinate (v1) at ([shift={(\Onex,\Oney,\Onez)}]v0);
\coordinate (vz) at ([shift={(\Zx,\Zy,\Zz)}]v0);
\coordinate (vi) at ([shift={(\tx,\ty,\tz)}]v0);

\filldraw[black!50] (0,0,0) circle (1.3pt);

\draw[dashed,-{Stealth[length=10pt, width=15pt]},black!50] (0,0,0) -- node[inner sep=1.5pt,below right] {$z_0$} (v0);
\draw[dashed,-{Stealth[length=10pt, width=15pt]},black!50] (0,0,0) -- node[inner sep=1.5pt,below left] {$z_1$} (v1);
\draw[dashed,-{Stealth[length=10pt, width=15pt]},black!50] (0,0,0) -- node[inner sep=1.5pt,below right] {$z_2$} (vz);
\draw[thin,->] ($(v0)!-\axisundershoot/\OneLen!(v1)$) -- ($(v0)!{1+2.7*\axisovershoot/\OneLen}!(v1)$) node[perspective,anchor=north west,opacity=1]{$\Re z$};
\draw[thin,->] ($(v0)!-\axisundershoot/\iLen!(vi)$) -- ($(v0)!{\OneLen/\iLen+1.4*\axisovershoot/\iLen}!(vi)$) node[perspective,anchor=south west,opacity=1]{$\Im z$};

\coordinate (C1) at ($(v0)!{2*\OneLen/\iLen}!(vi)$);
\node[perspective2] (C) at ($(C1)!{.5}!(v1)$) {$\CC$};

\pgfmathsetmacro{\xscale}{1.7};
\coordinate (R0) at (${1 + 2.1*\axisundershoot/\OneLen + 2*\axisundershoot/\iLen}*(v0) - 2.1*\axisundershoot/\OneLen*(v1)- 2*\axisundershoot/\iLen*(vi)$);
\coordinate (R1) at (${ - \xscale*2*\axisovershoot/\OneLen + 2*\axisundershoot/\iLen}*(v0) + {1 + 2*\xscale*\axisovershoot/\OneLen}*(v1)- 2*\axisundershoot/\iLen*(vi)$);
\coordinate (R2) at (${ - \OneLen/\iLen - 2*\xscale*\axisovershoot/\OneLen - 2*\axisovershoot/\iLen}*(v0) + {1 + 2*\xscale*\axisovershoot/\OneLen}*(v1)+ {\OneLen/\iLen + 2*\axisovershoot/\iLen}*(vi)$);
\coordinate (R3) at (${1 - \OneLen/\iLen + 2.1*\axisundershoot/\OneLen - 2*\axisovershoot/\iLen}*(v0) - {2.1*\axisundershoot/\OneLen}*(v1)+ {\OneLen/\iLen + 2*\axisovershoot/\iLen}*(vi)$);

\draw[opacity = .05,fill,black!50]  (R0) -- (R1) -- (R2) -- (R3);

\filldraw (v0) circle(1.3pt) node[below left,perspective] {$0$};
\filldraw (v1) circle(1.3pt) node[below,perspective] {$1$};
\filldraw (vz) circle(1.3pt) node[above right,perspective] {$z$};

\draw[thick] (v0) -- node[perspective,below]{$1$} (v1);
\draw[thick] (v0) -- node[perspective,above left]{$|z|$} (vz);
\draw[thick] (vz) -- node[perspective,below right]{$|z-1|$} (v1);

\end{tikzpicture}
\caption{The vectors $z_0,z_1,z_2\in\RR^D$ span a plane that is identified with the complex plane $\CC$ by requiring $z_0,z_1$ to coincide with $0,1\in \CC$.
The position of the vector $z_2$ inside the plane determines the value $z\in\CC$ (up to conjugation). Comparing ratios of squared side lengths of the triangles $z_0,z_1,z_2 \in \RR^D$
and $0,1,z \in \CC$ gives the relations in \eqref{eqinvs}.}
\label{fig:Ctriangle}
\end{figure}
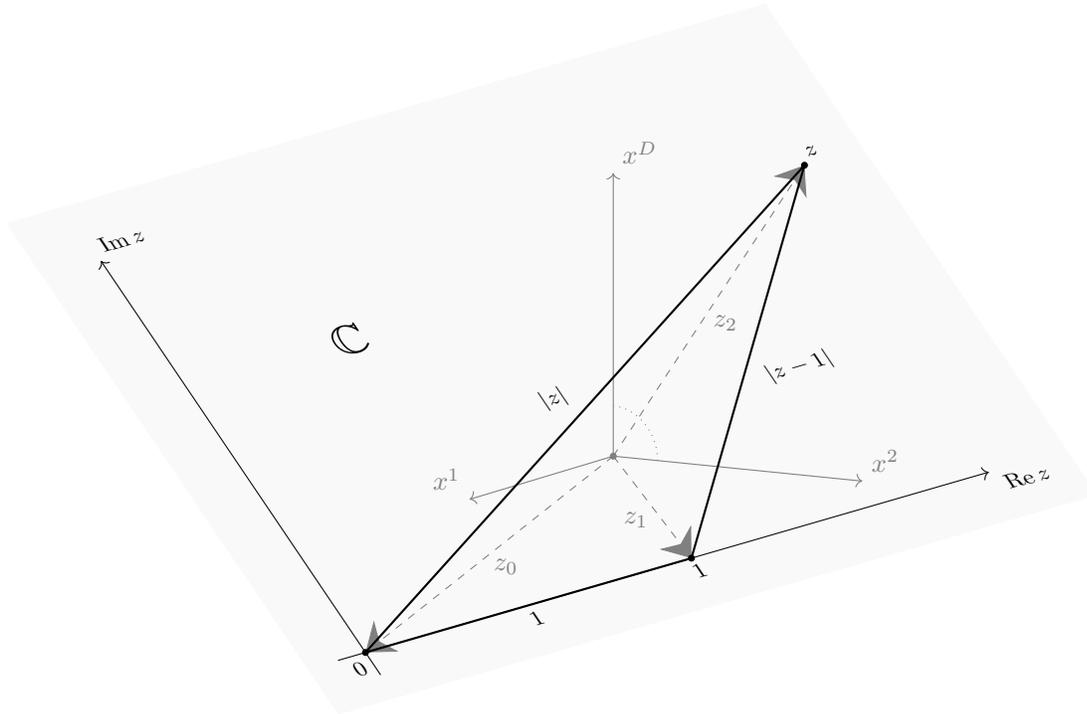

Graphical functions transfer the information of a position space Feynman integral $A_G(z_0,z_1,z_2)$ for a fixed graph $G$ with three external vertices $z_0$, $z_1$, $z_2$ to the complex plane.
The general idea is that the three external vertices span a two-dimensional subspace in $D$-dimensional ambient space. By symmetry, the Feynman integral in $\RR^D$ is
fully determined by its values on the two-dimensional subspace. We use invariance under translations to move the origin of the two-dimensional subspace to $z_0$ (say).
By rotational invariance we move the vector $z_1$ to the first coordinate axis. The (convergent) Feynman integral transforms by power-counting under a scale change. With this information we
scale $z_1$ to $(1,0,\ldots,0)$ and identify the two-dimensional subspace with the complex plane where $z_0$ corresponds to $0\in\CC$ and $z_1$ corresponds to $1\in\CC$.
The position of $z_2$ in the complex plane is fixed up to complex conjugation in $\CC$.
We fix one of the two choices for $z_2$ to define $z\in\CC$ as the sole variable on which the Feynman integral depends when restricted to $\CC$.
This gives a function $f_G(z)$ which is invariant under complex conjugation of the argument,
\begin{equation}\label{fGsym}
f_G(z)=f_G(\zz).
\end{equation}
The function $f_G(z)$ is called the graphical function of the graph $G$, see Figure \ref{fig:Ctriangle}.
If we want to emphasize the dimension $D$ of ambient space we write $f_G^{(\lambda)}(z)$ for $\lambda=D/2-1$.

Graphical functions were first defined in \cite{gf} with the aim to calculate the residues of four-dimensional Feynman integrals without subdivergences (Feynman periods).
In collaboration with Michael Borinsky, the author was recently able to generalize the theory of graphical functions to all even dimensions $\geq4$ \cite{gfe}.
The concept of graphical functions in four dimensions also (independently) surfaces as conformal integrals in $N=4$ super Yang-Mills theory (see e.g.\ \cite{SYM}).

In this article we generalize graphical functions to $D=2n+4-\epsilon$ `dimensions' ($n=0,1,2,\ldots$)\footnote{Many formulas are somewhat simpler with $D=2n+4-2\epsilon$ but for
consistency with some literature and with \cite{Shlog} we use $D=2n+4-\epsilon$.}, where the dimension serves as a parameter which regularizes integrals with divergences.
This commonly used technique of dimensional regularization is particularly suited for using graphical functions. One assumes that $\epsilon$ is a small parameter and
aims to calculate the Laurent expansion of Feynman integrals up to a certain order in $\epsilon$. In this context (and not only in this context) it is useful to adopt a second (equivalent)
definition of graphical functions in terms of invariants. We note that from the three external vertices $z_0,z_1,z_2$ we can build the two scalar invariants (see Figure \ref{fig:Ctriangle})
\begin{equation}\label{eqinvs}
\frac{\|z_2-z_0\|^2}{\|z_1-z_0\|^2}=z\zz,\quad\frac{\|z_2-z_1\|^2}{\|z_1-z_0\|^2}=(z-1)(\zz-1).
\end{equation}
With these invariants we can express $A_G(z_0,z_1,z_2)$ in terms of $z$ (and $\zz$),
\begin{equation}\label{AGinvs}
A_G(z_0,z_1,z_2)=||z_1-z_0||^{-2\lambda N_G}f_G(z),
\end{equation}
where we used
\begin{equation}\label{NGdef}
N_G=\Big(\sum_e\nu_e\Big)-\frac{\lambda+1}{\lambda}V_G^{\mathrm{int}}.
\end{equation}
In $N_G$ the sum is over all edges of $G$, and $V_G^{\mathrm{int}}$ is the number of internal vertices of $G$ (i.e.\ vertices $\neq z_0,z_1,z_2$).
The weight $N_G$ is closely related to the `superficial degree of divergence' of $G$.

The connection to $z\in\CC$ in (\ref{eqinvs}) is evident from the correspondence $z_0,z_1,z_2\mapsto0,1,z$. The setup is equivalent to calculating a conformally invariant four-point function,
where (\ref{eqinvs}) is replaced by the cross ratios of the four external vertices $z_0,z_1,z_2,z_3$, with $z_3$ as the point at infinity, see e.g.\ \cite{SYM}.
The map from three-point graphical functions (with no conformal invariance) to conformal four-point functions is called completion in the theory of graphical functions, see Section \ref{sectcom}.
For a definition of graphical functions in non-integer dimensions one has to use the invariants in (\ref{eqinvs}) in combination with a parametric representation of graphical functions
\cite{par,gfe}. For practical purposes, however, the parametric representation is often unwieldy. One rather uses identities which are derived in general integer dimensions while interpreting
the dimension as a real parameter. This mathematically not concise approach has been tested and justified by the contradiction-free calculation of a huge number of graphical functions with the
Maple package {\tt HyperlogProcedures} \cite{Shlog}.

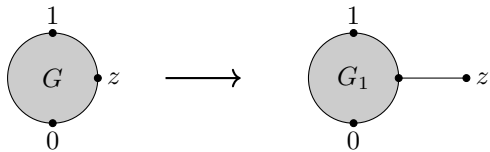
\begin{figure}
\begin{align*}
    \def\rad{.6}
\begin{tikzpicture}[baseline={([yshift=-.7ex](0,0))}]
    \coordinate (v) at (0,0);
    \draw[fill=black!20] (v) circle (\rad);
    \node (G) at (v) {$G$};
    \coordinate[label=above:$1$] (v1) at ([shift=(90:\rad)]v);
    \coordinate[label=below:$0$] (v0) at ([shift=(-90:\rad)]v);
    \coordinate[label=right:$z$] (vz) at ([shift=(0:\rad)]v);
    \coordinate (vm) at ([shift=(0:1.5)]v);
    \filldraw (v0) circle (1.3pt);
    \filldraw (v1) circle (1.3pt);
    \filldraw (vz) circle (1.3pt);
    \coordinate (w) at (4,0);
    \draw[fill=black!20] (w) circle (\rad);
    \node (G1) at (w) {$G_1$};
    \coordinate[label=above:$1$] (w1) at ([shift=(90:\rad)]w);
    \coordinate[label=below:$0$] (w0) at ([shift=(-90:\rad)]w);
    \coordinate (wx) at ([shift=(0:\rad)]w);
    \coordinate[label=right:$z$] (wz) at ([shift=(0:{2.5*\rad})]w);
    \coordinate (wm) at ([shift=(180:1.5)]w);
    \draw (wx) -- (wz);
    \draw[thick,->] (vm) -- (wm);
    \filldraw (w0) circle (1.3pt);
    \filldraw (w1) circle (1.3pt);
    \filldraw (wx) circle (1.3pt);
    \filldraw (wz) circle (1.3pt);
\end{tikzpicture}
\end{align*}
\label{fig:append}
\caption{By appending an edge to $z$ we obtain the graphical function $f_{G_1}(z)$ from the simpler graphical function $f_G(z)$.}
\end{figure}

The benefit of using graphical functions is their relative simplicity (compared to general Feynman integrals). Using complex numbers often linearizes denominators: We see in (\ref{eqinvs})
that the right hand sides factorize into linear factors whereas the left hand sides do not. This effect combines with the much deeper (and more surprising) result that it is always possible
to append an edge to an external vertex of a known graphical function, see Figure \ref{fig:append}.
There exists a complicated but fully explicit algorithm to calculate the more complex graphical function $f_{G_1}(z)$ from the simpler $f_G(z)$ by a series by single-valued integrations \cite{gfe}.
In many cases the algorithms is surprisingly efficient \cite{Shlog}. A key result in this article is that this algorithm extends to $2n+4-\epsilon$ dimensions.

In practice, one is often interested in a scalar process which emerges from a two-point Feynman integral. In particular, the calculation of renormalization functions is best done by
constructing a two-point limit of the four-point amplitude in $\phi^4$ (or the three-point amplitude in $\phi^3$), see Section \ref{sectren}.
For calculating scalar processes, the graphical function method introduces a third external vertex $z_2$ in a suitable way.
At the end of the calculation $z_2$ is eliminated by setting $z_2=z_0$ (or $z_2=z_1$), see Section \ref{sectper}. With the extra variable $z_2$ one obtains a richer theory which helps to
compute the scalar process \cite{numfunct}. On the other hand one does not want to handle two or more extra external vertices (making the theory even richer) because
this would make the (intermediate) Feynman integrals too complicated.

In general, graphical functions are defined in Euclidean space. Because of the connection to the invariants (\ref{eqinvs}) one can equivalently consider a Lorentzian metric by using
$z$ and $\zz$ as independent real variables (instead of complex conjugates). A detailed survey of the theory of graphical functions in even dimensions is \cite{gfe}.
The connection between graphical functions and number theory with the seven loop results for the $\phi^4$ renormalization functions is in \cite{numfunct}.
A short and more physical account of graphical functions in $2n+4-\epsilon$ dimensions is in Section 3.1 of \cite{5lphi3}.

Here, we restrict ourselves to the definition and the basic properties of graphical functions in $2n+4-\epsilon$ dimensions. We define $\lambda=n+1-\epsilon/2$ so that
\begin{equation}\label{Ddef}
D=2\lambda+2=2n+4-\epsilon,\quad n=0,1,2,\ldots.
\end{equation}

It is very convenient to weight the edges of the graph $G$. The position space propagator of the edge from $x$ to $y$ with weight $\nu$ is
\begin{equation}\label{propagator}
p_{xy}=||x-y||^{-2\nu\lambda},
\end{equation}
where $||x-y||$ is the Euclidean distance between $x$ and $y$ in $D$ dimensions. We integrate over all internal vertices where
each $D$-dimensional integral is normalized by $\pi^{-D/2}$. This gives the dimensionally regularized Feynman integral $A_G(z_0,z_1,z_2)$.

The identification of the two-dimensional plane spanned by $z_0,z_1,z_2$ with $\CC$ is explicit if we choose the coordinates
\begin{equation}\label{eqzdef}
z_0=0=\left(\begin{array}{c}0\\0\\0\\ \vdots\\0\end{array}\right),\quad
z_1=\left(\begin{array}{c}1\\0\\0\\\vdots\\0\end{array}\right),\quad
z_2(z)=\left(\begin{array}{c}\Re z\\\Im z\\0\\\vdots\\0\end{array}\right).
\end{equation}
In these coordinates we get from (\ref{AGinvs}),
$$
f_G(z)=A_G(0,z_1,z_2(z)).
$$

In general, $f_G(z)$ admits a Laurent expansion in $\epsilon$ whose coefficients inherit the three general properties of graphical functions in even integer dimensions.
General property (G1) is the symmetry under complex conjugation (\ref{fGsym}). General property (G2) is real-analyticity combined with single-valuedness. It is proved
in \cite{par} for all dimensions $D$ that $f_G(z)$ is a single-valued real-analytic function on $\CC\backslash\{0,1\}$. So, every Laurent-coefficient of $f_G(z)$
has this property. The general property (G3) is stronger than (G2). It says that every Laurent coefficient admits single-valued log-Laurent expansions at the singular
points $0,1,\infty$. Explicitly, for $s=0,1$ we have
\begin{equation}\label{01expansion}
\text{$\epsilon^k$ coefficient of }f_G(z)=\sum_{\ell\geq0}\sum_{m,\mm=M_s}^\infty c_{\ell,m,\mm}^{s,k}[\log(z-s)(\zz-s)]^\ell(z-s)^m(\zz-s)^\mm\quad\text{if }|z-s|<1
\end{equation}
for some coefficients $c_{\ell,m,\mm}^{s,k}\in\RR$. At infinity we have the expansion
\begin{equation}\label{inftyexpansion}
\text{$\epsilon^k$ coefficient of }f_G(z)=\sum_{\ell\geq0}\sum_{m,\mm=-\infty}^{M_\infty}c_{\ell,m,\mm}^{\infty,k}(\log z\zz)^\ell z^m\zz^\mm\quad\text{if }|z|>1
\end{equation}
for constants $c_{\ell,m,\mm}^{\infty,k}\in\RR$. The existence of single-valued log-Laurent expansions at the singular points is an important technical condition.
Moreover, for calculations with graphical functions one needs to restrict to a good function space. The space of generalized single-valued hyperlogarithms (GSVHs) in \cite{GSVH} is
particularly suited for representing graphical functions. The construction of GSVHs intrinsically relies on the existence of single-valued log-Laurent expansions.

\section{Renormalization}\label{sectren}
For the calculations presented in this article we use a simplistic approach to renormalization. For each Feynman graph we calculate the $\epsilon$-expansion of the Feynman integral to
the needed order and extract the $Z$ factors $Z_\phi,Z_g,Z_m$ which scale the bare values of the field $\phi$, the coupling $g$, and the mass $m$, see e.g.\ \cite{IZ}.
There exist much more refined techniques for calculating renormalization functions in momentum space (such as infrared rearrangement in the $R^\ast$ operation). These techniques
can also be used in position space. However, these techniques are to some extent built-in features of the graphical functions method.
Still, a future calculation of the $\phi^4$ beta function to eight loops may benefit from the use of more refined techniques. Because of our naive approach we refrain from giving a more detailed
review on renormalization and refer to any text book in the field (see e.g.\ \cite{IZ}).

For the anomalous dimension $\gamma$ we only consider the two-point function. We restrict ourselves to one particle irreducible (1PI) graphs.
For the $\ell$ loop result we need to calculate 1PI graphs of loop orders $k\leq\ell$ to order $\epsilon^{\ell-k-1}$. We determine the Laurent expansions of $Z_\phi,Z_g,Z_m$ such
that the result is regular for $\epsilon\to0$.
In the wake of calculating the eight loop result for the $\phi^4$ gamma function (\ref{8loopgamma}) we are able to extract the self-energy to loop order seven,

\begin{align}
\frac{\Sigma_7^{\phi^4}(p)}{p^2}&=\frac{81}{2}L^6+\frac{5721}{10}L^5+\left(\frac{352745}{96}+90\zeta(3)\right)L^4+\left(\frac{437223}{32}+\frac{2074}{3}\zeta(3)+\frac{1}{5}\pi^4+180\zeta(5)\right)L^3\nonumber\\
&+\,\left(\frac{47340787}{1536}+\frac{151703}{64}\zeta(3)+\frac{2047}{1440}\pi^4+\frac{3817}{4}\zeta(5)+\frac{5}{84}\pi^6-\frac{3}{2}\zeta(3)^2+\frac{441}{2}\zeta(7)\right)L^2\nonumber\\
&+\,\left(\frac{912929119}{23040}+\frac{2736521}{640}\zeta(3)+\frac{26803}{7200}\pi^4+\frac{15351}{8}\zeta(5)+\frac{278}{945}\pi^6+\frac{1543}{20}\zeta(3)^2+\frac{1}{20}\pi^4\zeta(3)\right.\nonumber\\
&\quad\left.+\,\frac{20919}{40}\zeta(7)+\frac{2063}{105000}\pi^8-54\zeta(3)\zeta(5)-\frac{324}{25}\zeta(5,3)\right)L\nonumber\\
&+\,\frac{11539080223}{516096}+\frac{71852411}{21504}\zeta(3)+\frac{2793299}{806400}\pi^4+\frac{7472403}{4480}\zeta(5)+\frac{431191}{1270080}\pi^6+\frac{1086739}{6720}\zeta(3)^2\nonumber\\
&\quad+\,\frac{2201}{16800}\pi^4\zeta(3)+\frac{150337}{448}\zeta(7)+\frac{134413}{5880000}\pi^8+\frac{66}{7}\zeta(3)\zeta(5)-\frac{261}{350}\zeta(5,3)-\frac{1}{98}\pi^6\zeta(3)
\nonumber\\
&\quad+\,\frac{9}{700}\pi^4\zeta(5)-\frac{79}{35}\zeta(3)^3+\frac{16421}{2520}\zeta(9),
\end{align}
where
$$
L=\frac{1}{2}\log\left(\frac{4\pi\Lambda^2}{\ee^Cp^2}\right),
$$
with the renormalization scale $\Lambda$ and Euler-Mascheroni constant $C=0.577\ldots$.
The result for the first six loops is in \cite{numfunct}.

The five loop result for the self-energy in six-dimensional $\phi^3$ theory is
\begin{align}
\frac{\Sigma_5^{\phi^3}(p)}{p^2}&=\left(-\frac{665}{3888}L^4-\frac{33985}{23328}L^3+\left(-\frac{916571}{186624}+\frac{7}{72}\zeta(3)\right)L^2\right.\nonumber\\
&\quad+\,\left(-\frac{4276967}{559872}-\frac{413}{3888}\zeta(3)-\frac{7}{12960}\pi^4+\frac{5}{9}\zeta(5)\right)L\nonumber\\
&\left.\quad-\,\frac{39457147}{8957952}-\frac{55093}{93312}\zeta(3)-\frac{553}{622080}\pi^4+\frac{193}{288}\zeta(5)+\frac{5}{27216}\pi^6-\frac{1}{72}\zeta(3)^2\right)g^8\nonumber\\
&+\,\left(\frac{95}{648}L^3+\frac{2155}{2592}L^2+\left(\frac{25801}{15552}-\frac{1}{12}\zeta(3)\right)L+\frac{419449}{373248}-\frac{65}{1296}\zeta(3)-\frac{1}{4320}\pi^4\right)g^6\nonumber\\
&+\,\left(-\frac{5}{36}L^2-\frac{31}{72}L-\frac{1789}{5184}\right)g^4\;+\;\left(\frac{1}{6}L+\frac{2}{9}\right)g^2.
\end{align}

With the $\ell$ loop results for the 1PI two-point function one can also determine the beta function and the mass anomalous dimension $\gamma_m$ to loop order $\ell-2$.
For higher loop orders it is convenient to consider the four-point function in $\phi^4$ theory and the three-point function in $\phi^3$ theory. Again, we can restrict ourselves to 1PI processes.
It is inefficient to calculate the full $N$-point function for extracting the $Z$-factors. We rather reduce the $N$-point function to an effective two-point function by taking limits.
We truncate one external leg by considering the limit $z_N\to\infty$ of $|z_N|^{2\lambda}A_G(\ldots,z_N)$ where $A_G$ is the Feynman integral for the 1PI graph $G$ and $z_N$ is one
of its external vertices. Because $G$ is 1PI, the limit $z_N\to\infty$ coincides with the naive limit taken in the integrand (this limit truncates the leg attached to $z_N$):
Other potential terms arise after scaling (blowing up in mathematical terminology) a subset of integration variables by $x_i\mapsto|z_N|x_i$. Such a scaling generates a
weight $|z_N|^{DV_g^{\mathrm{int}}-2\lambda E_g}$, where $V_g^{\mathrm{int}}$ and $E_g$ are the number of internal (scaled) vertices and the number of edges of the subgraph $g$.
The subgraph $g$ has all edges of $G$ that are attached to at least one scaled vertex.
By counting half-edges we get $2E_g-N_g=V_g^{\mathrm{int}}D/\lambda+\sO(\epsilon)$, where $N_g$ is the number of external legs of $g$.
We find a total scaling behavior $|z_N|^{-N_g\lambda+\sO(\epsilon)}$ with an extra factor $|z_N|^{-2\lambda}$ if the edge attached to $z_N$ is not in $g$.
For a non-trivial subgraph $g$ in a 1PI graph we get $N_g\geq3$, so that these terms are suppressed compared to the trivial case $g=\emptyset$ (with the scaling weight $|z_N|^{-2\lambda}$).

If we consider the three-point function in $\phi^3$ theory, we obtain the desired two-point function after sending one vertex to infinity. In $\phi^4$ theory we are left with an effective
three-point function. This three-point function has a three-valent vertex which alters the above counting argument. In fact, we cannot take the limit $z_{N-1}\to\infty$ by
truncating the leg that is attached to $z_{N-1}$. In general, there may exist several terms that contribute to the limit.

\begin{figure}
\centering
\includegraphics{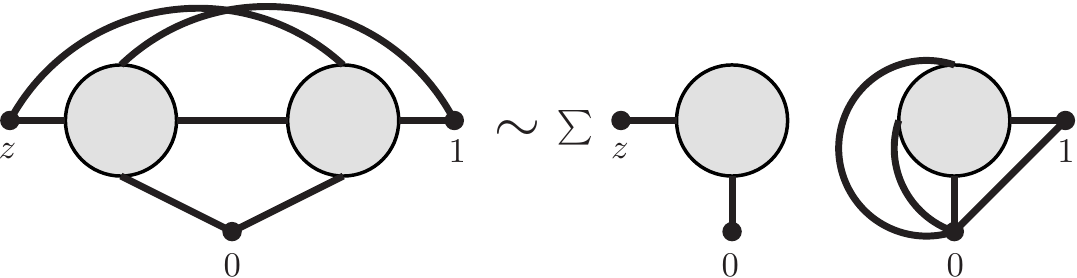}
\caption{The asymptotic expansion of graphical functions at $z=0$. Bold lines stand for sets of edges.}
\label{fig:asympt}
\end{figure}

Because the effective three-point function may be considered as a graphical function, the situation is within the general setup of calculating limits $z\to0,1,\infty$ in graphical functions.
It does not matter if we take the limit $z\to0$, $z\to1$, or $z\to\infty$ because permutations of external vertices permute the singularities $0$, $1$, $\infty$ (Section \ref{sectcon}) and,
in the end, we sum over all permutations of external vertices. The limits are calculated according to the scaling method described above. The result for the limit $z\to 0$ is
illustrated in Figure \ref{fig:asympt}, where on the right hand side one has to sum over all partitions of internal vertices into two disjoint sets \cite{numfunct}.

The general statement (mathematically a well-tested conjecture) is as follows.
Assume $G$ is a graph such that the graphical function $f_G(z)$ exists. Let $\sV^{\mathrm{int}}_G$ and $\sV^{\mathrm{ext}}_G=\{0,1,z\}$ be the sets of internal and external vertices of $G$,
respectively.  For $V\subseteq\sV^{\mathrm{int}}_G$ let $G[V]$ be the subgraph of $G$ which is induced by $V$, i.e.\ the subgraph which contains the vertices $V$ and all edges of $G$ with both
vertices in $V$. Further let $G[V=0]$ be the contracted graph $G/G[V]$ where one identifies all vertices in $V$ with the vertex $0$. Then, we obtain the asymptotic expansions at $z=0$ by
\begin{equation}\label{0asympt}
f_G(z)=\sum_{V\subseteq\sV^{\mathrm{int}}_G}f_{G[V\cup\{0,z\}]}(z)f_{G[V\cup\{0,z\}=0]}(1+O(|z|^2))
\end{equation}
whenever the right hand side exists. Note that the right hand side of (\ref{0asympt}) only has two-point graphs.

The result is not symmetric under permutations of external vertices. This leads to a proliferation of graphs as we have to deal with leg-fixed graphs where we
distinguish graphs which have different labels at external legs. The resulting two-point functions have a residual reflection symmetry, so that the total proliferation
factor is $4!/2=12$ for $\phi^4$ theory and $3!/2=3$ for $\phi^3$ theory. In practice, this is not as severe as it looks because in most cases the calculations for permuted external vertices
are either identical or very similar. The program {\tt HyperlogProcedures} memorizes previous results with the effect that results are often instantaneous if a permuted contribution
has been calculated beforehand.

In $\phi^4$ theory we extract the $Z$-factors for the beta function from the coefficients of $\log(z\zz)$ in the effective two-point function. At top loop order we need the
coefficient $\epsilon^{-2}$. This is equivalent to calculating the order $\epsilon^{-1}$ of the amputated graph (where the external legs are removed) in momentum space.

For loop orders $k\leq\ell$ we hence calculate the $\epsilon$-expansion to order $\epsilon^{\ell-k-2}$ if we are interested in the $\ell$ loop result for
the beta function. The advantage of extracting the $Z$-factor from the coefficients of $\log(z\zz)$ is that we can skip the calculation of the terms which are constant in $z$.
These are the terms with $z=0$ in the integrand (which can be quite hard to compute). In $\phi^3$ theory such a trick is not possible; to compute the $\ell$ loop beta function
one has to expand $k\leq\ell$ loop graphs to order $\epsilon^{\ell-k-1}$.

\section{Appending an edge}\label{sectappedge}
We aim to calculate graphical functions by single-valued integration. This is facilitated by an algorithm that appends an edge of weight 1 to the external vertex $z$, see Figure \ref{fig:append}
\cite{gf,gfe,numfunct,5lphi3}. The many identities between graphical functions, see Sections \ref{sectid} -- \ref{sectrerou}, have the sole purpose to reduce the calculation of
a graphical function where no edge can be appended to a sum of graphical functions which are amenable to this reduction.

Although we have introduced the regulator $\epsilon$ in (\ref{Ddef}), in this section we only consider the case that a graphical function is convergent in the limit $\epsilon\to0$.
For the appended graph we will obtain a power series in $\epsilon$ whose coefficients can be calculated. In QFT, Feynman integrals normally have divergences which can be handled
with a subtraction procedure, see Section \ref{sectreg}.

The propagator attached to the external vertex $z$ is the Green's function of the $D$-dimensional Laplacian. For the corresponding graphical function this translates into
an effective Laplace equation (\cite{gf} with a proof in non-integer dimensions in \cite{GolzMaster}),
\begin{gather}
\begin{gathered}
\label{eq:effLap}
\left(\Delta_n+\frac{\varepsilon/2}{z-\zz}(\partial_{z} - \partial_{\zz})\right)
\left[
\begin{tikzpicture}[x=1.5ex,y=1.5ex,baseline={([yshift=-.6ex]current bounding box.center)}]
    \coordinate (v0);
    \coordinate [above right=1.41 and 2 of v0] (v1);
    \coordinate [below right=1.41 and 2 of v0] (v2);
    \coordinate [left =3 of v0] (v00);
    \node [left =.1 of v00] (v0n) { $z$};
    \coordinate [right =1 of v0] (v0u);
    \node [right=.07 of v1] (v1n) { $1$};
    \coordinate [below left=.71 and .71 of v1] (v1u);
    \node [right=.07 of v2] (v2n) { $0$};
    \coordinate [above left=.71 and .71 of v2] (v2u);
    \draw[fill=gray] (v0) .. controls (v0u) and (v1u) .. (v1) .. controls (v1u) and (v2u) .. (v2) .. controls (v2u) and (v0u) .. (v0);
    \draw (v00) -- (v0);
    \filldraw (v00) circle (1pt);
    \filldraw (v0) circle (1pt);
    \filldraw (v1) circle (1pt);
    \filldraw (v2) circle (1pt);
\end{tikzpicture}%
\right]
=-\frac{1}{\Gamma(\lambda)}
\left[
\begin{tikzpicture}[x=1.5ex,y=1.5ex,baseline={([yshift=-.6ex]current bounding box.center)}]
    \coordinate (v0);
    \coordinate [above right=1.41 and 2 of v0] (v1);
    \coordinate [below right=1.41 and 2 of v0] (v2);
    \node [left =.1 of v0] (v0n) { $z$};
    \coordinate [right =1 of v0] (v0u);
    \node [right=.07 of v1] (v1n) { $1$};
    \coordinate [below left=.71 and .71 of v1] (v1u);
    \node [right=.07 of v2] (v2n) { $0$};
    \coordinate [above left=.71 and .71 of v2] (v2u);
    \draw[fill=gray] (v0) .. controls (v0u) and (v1u) .. (v1) .. controls (v1u) and (v2u) .. (v2) .. controls (v2u) and (v0u) .. (v0);
    \filldraw (v0) circle (1pt);
    \filldraw (v1) circle (1pt);
    \filldraw (v2) circle (1pt);
\end{tikzpicture}%
\right]
\\
\text{with}\quad
\Delta_n = \frac{1}{(z-\zz)^{n+1}}\partial_{z}\partial_{\zz}(z-\zz)^{n+1}+\frac{n(n+1)}{(z-\zz)^2}.
\end{gathered}
\end{gather}
Here, $\Gamma(x)=\int_0^\infty t^{x-1}\exp(-t)\dd t$ is the gamma function. For convergent graphical functions and $\epsilon=0$ the inversion of $\Delta_n$ is solved in \cite{gfe}.
The algorithm demands the ability to calculate single-valued \mbox{(anti-)}primitives. The space of generalized single-valued hyperlogarithms (GSVHs) suffices for the calculation of
loop orders $\leq7$ in $\phi^4$ theory and (at least) $\leq6$ in $\phi^3$ theory. There exists a fully proved theory of GSVHs which provides efficient
algorithms for single-valued integration \cite{GSVH}. At high loop orders the Feynman integrals of some graphs lead out of the space of GSVHs. This
has prominently been proved by the existence of a K3 surface in $\phi^4$ theory at eight loops \cite{K3}. The period of the K3 surface is one of the non-hyperlogarithmic contributions
to the eight loop beta function in $\phi^4$ theory. An account of geometries which are to be expected in QFT calculations can be obtained via the $c_2$-invariant, see \cite{SchnetzFq,Sc2}
(and the references therein). See also \cite{motg2} for the number content of the electron anomalous magnetic moment. We expect that only four primitive (subdivergence-free) graphs
give non-hyperlogarithmic contributions to the eight loops $\phi^4$ beta function. In particular, we expect that all subdivergent eight loop graphs in $\phi^4$ theory are hyperlogarithmic.
The contributions of the four non-hyperlogarithmic graphs at eight loops can be calculated numerically with the tropical Monte Carlo Method by M. Borinsky \cite{TMCQ}. 

In alignment with Figure \ref{fig:append} we denote the graphs on the left and right hand sides of (\ref{eq:effLap}) by $G_1$ and $G$, respectively.
To solve (\ref{eq:effLap}) to order $\epsilon^n$ we expand the graphical function $f_G(z)$ in powers of $\epsilon$,
\begin{equation}\label{fGexp}
f_G(z)=f_{G,0}(z)+\epsilon f_{G,1}(z)+\epsilon^2 f_{G,2}(z)+\ldots+\epsilon^n f_{G,n}(z).
\end{equation}
With this ansatz for $f_G(z)$ we iteratively solve (\ref{eq:effLap}) order by order in $\epsilon$. We invert $\Delta_n$ and obtain
\begin{equation}\label{fGexp1}
\Big(1+\Delta_n^{-1}\frac{\epsilon/2}{z-\zz}(\partial_z-\partial_\zz)\Big)f_{G_1}(z)=-\frac{1}{\Gamma(\lambda)}\Delta_n^{-1}f_G(z).
\end{equation}
The inverse of $\Delta_n$ is proved to be unique for convergent four-dimensional graphical functions and $\epsilon=0$ \cite{gfe}. It is a very well tested conjecture that uniqueness
is true for all orders in $\epsilon$ in any even dimension $\geq4$. Uniqueness, however, is lost for graphical functions which are singular at $\epsilon=0$ (although these
graphical functions exist for $\epsilon\neq0$). We need a subtraction procedure to regularize graphical functions which are divergent in integer dimensions before we can
calculate their $\epsilon$-expansions. This procedure is detailed in the Section \ref{sectreg}. We emphasize that the subtraction procedure is not necessary to define graphical
functions for $\epsilon\neq0$ (where $\epsilon$ serves as regulator). The sole reason for the subtraction is to be able to use the algorithm in this section in order to calculate their
$\epsilon$-expansions.

The next step in the calculation of the graphical function $f_{G_1}(z)$ is to invert the operator on the left hand side of (\ref{fGexp1}) to power $\epsilon^n$ by
expanding the geometric series $(1+\Delta_n^{-1}\frac{\epsilon/2}{z-\zz}(\partial_z-\partial_\zz))^{-1}$. This yields
\begin{equation}\label{eqappend}
f_{G_1}(z)=-\frac{1}{\Gamma(\lambda)}\sum_{k=0}^n\Big(-\Delta_n^{-1}\frac{\epsilon/2}{z-\zz}(\partial_z-\partial_\zz)\Big)^k\Delta_n^{-1}f_G(z).
\end{equation}
The sum over $k$ is calculated term by term from $0$ to $n$. Every new term $k$ is obtained by inverting $\Delta_n$ on a derivative of term $k-1$ which is effectively truncated to order
$\epsilon^{n-k}$ (the derivative of term $k-1$ has low degree $k$ in $\epsilon$).

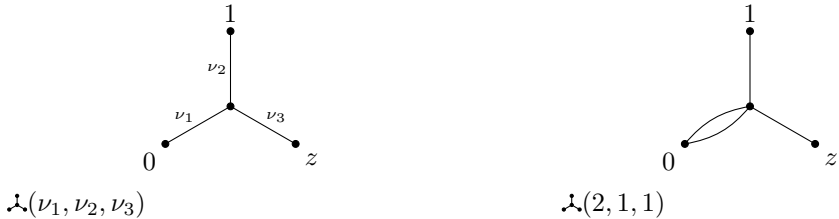
\begin{figure}
\begin{align*}
& 
\begin{tikzpicture}
    \coordinate (v) ;
    \def \rad {1};
    \coordinate[label=above:$1$] (v1) at ([shift=(90:\rad)]v);
    \coordinate[label=below left:$0$] (v2) at ([shift=(210:\rad)]v);
    \coordinate[label=below right:$z$] (v3) at ([shift=(330:\rad)]v);
    \draw (v) -- node[inner sep=1pt,left] {\tiny$\nu_2$} (v1);
    \draw (v) -- node[inner sep=1pt,above left] {\tiny$\nu_1$} (v2);
    \draw (v) -- node[inner sep=1pt,above right] {\tiny$\nu_3$} (v3);
    \filldraw (v) circle (1.3pt);
    \filldraw (v1) circle (1.3pt);
    \filldraw (v2) circle (1.3pt);
    \filldraw (v3) circle (1.3pt);
    \node [below left=of v] {$\smallclaw(\nu_1,\nu_2,\nu_3)$};
\end{tikzpicture}
\hspace{3cm}
\begin{tikzpicture}
    \coordinate (v) ;
    \def \rad {1};
    \coordinate[label=above:$1$] (v1) at ([shift=(90:\rad)]v);
    \coordinate[label=below left:$0$] (v2) at ([shift=(210:\rad)]v);
    \coordinate[label=below right:$z$] (v3) at ([shift=(330:\rad)]v);
    \draw (v) -- node[inner sep=1pt,left] {} (v1);
    \draw (v) to [bend left =20] (v2);
    \draw (v) to [bend right =20] (v2);
    \draw (v) -- node[inner sep=1pt,above right] {} (v3);
    \filldraw (v) circle (1.3pt);
    \filldraw (v1) circle (1.3pt);
    \filldraw (v2) circle (1.3pt);
    \filldraw (v3) circle (1.3pt);
    \node [below left=of v] {$\smallclaw(2,1,1)$};
\end{tikzpicture}
\end{align*}
\caption{
The weighted three-star $\smallclaw(\nu_1,\nu_2,\nu_3)$. The three-star one the right hand side has a bubble divergence in four dimensions.
}
\label{fig:3star}
\end{figure}

As an example we consider the three-star $\smallclaw(1,1,1)$ on the left hand side of Figure \ref{fig:3star} in $D=4-\epsilon$ dimensions (i.e.\ $n=0$).
The graphical function on the right hand side of (\ref{eq:effLap}) is $(z\zz)^{-\lambda}((z-1)(\zz-1))^{-\lambda}$, see (\ref{eqinvs}), (\ref{AGinvs}), and (\ref{propagator}).

For the term $k=0$ in (\ref{eqappend}) we obtain to order $\epsilon$ \cite{Shlog}
\begin{equation}\label{eqex1}
f_{G_1,0}(z)=\frac{(2-\epsilon\gamma)(\sL_{01}-\sL_{10})+\epsilon(\sL_{001}+\sL_{011}-\sL_{100}-\sL_{110})}{2(z-\zz)},
\end{equation}
where the single-valued hyperlogarithms $\sL_w(z)$ are inductively defined by single-valued integration,
$$
\sL_{wa}(z)=\intsv\frac{\sL_w(z)}{z-a}\dd z\qquad\text{and}\qquad\sL_e(z)=1
$$
for the empty word $e$ \cite{BrSVMP,GSVH}. Note that $(\sL_{01}-\sL_{10})/4\ii$ is the Bloch-Wigner dilogarithm \cite{Zagierdilog}.

The function $f_{G_1,0}(z)$ in (\ref{eqex1}) gives the order $\epsilon^0$ of $f_{G_1}(z)$.
The differential operator $\frac{\epsilon/2}{z-\zz}(\partial_z-\partial_\zz)$ on $f_{G_1,0}(z)$ yields
$$
\epsilon\Big(\frac{\sL_{01}-\sL_{10}}{(z-\zz)^3}+\frac{(2-z-\zz)\sL_0}{2(z-\zz)^2(z-1)(\zz-1)}+\frac{(z+\zz)\sL_1}{2(z-\zz)^2z\zz}\Big)+\sO(\epsilon^2).
$$
Inverting $\Delta_n$, this integrates to
\begin{equation}\label{eqex2}
\epsilon f_{G_1,1}(z)=\epsilon\frac{-\sL_{010}+\sL_{101}+2\sL_{01\zz}-2\sL_{10\zz}}{2(z-\zz)}+\sO(\epsilon^2).
\end{equation}
The expression $\sL_{01\zz}-\sL_{10\zz}$ is a GSVH which is not a single-valued hyperlogarithm. It is the first GSVH that was studied in QFT \cite{Duhr}.
Dimensionally regularized graphical functions very frequently use the alphabet $0,1,\zz$, see Section 8.3 in \cite{GSVH}.

We add $f_{G_1,0}(z)$ in (\ref{eqex1}) and $\epsilon f_{G_1,1}(z)$ in (\ref{eqex2}) to obtain
$$
f_{G_1}(z)=f_{G_1,0}(z)+\epsilon f_{G_1,1}(z)+\sO(\epsilon^2).
$$
With {\tt HyperlogProcedures} \cite{Shlog} running on an office PC one can calculate $\smallclaw(1,1,1)$ to order $\epsilon^{10}$.

\section{Subtraction of subdivergences}\label{sectreg}
Feynman integrals in QFT have divergences. In four-dimensional $\phi^4$ theory a frequent case is the bubble which is depicted as the edge attached to 0 in the three-star $\smallclaw(2,1,1)$
in Figure \ref{fig:3star}. Although divergences are regularized by the transition to non-integer dimensions, we cannot directly use the algorithm of appending an edge in Section \ref{sectappedge}.
In presence of a divergence (and only in this case) the result for the graphical function with appended edge has a term which is in the kernel of the effective Laplace equation
(\ref{eq:effLap}). The algorithm for appending an edge in the final section of \cite{gfe} is blind to this kernel and gives a false result.

The solution of this problem is a subtraction procedure. The subtraction kills the term which is in the kernel of (\ref{eq:effLap}). In general, it suffices to handle the subtraction of
logarithmic subdivergences. (It is possible to subtract poles of higher order with a similar but more tedious method.) The only case in $\phi^4$ theory with divergences of higher
order are two-point insertions which can trivially be factored out, see Sections \ref{sectper} and \ref{sectcon}.

A logarithmic divergence arises from a pole of order $2n+4$ at $z=0$ or at $z=1$ in the graphical function $f_G(z)$ for $\epsilon=0$.
Using completion in Section \ref{sectid} can lead to an infrared divergence. This corresponds to the point $z=\infty$ in $f_G(z)$, where a logarithmic divergence
is of order $\sO(|z|^{-2})$ for $z\to\infty$.

We first consider the case of a pole of order $2n+4$ at $z=0$. The method is illustrated in the case of the three-star $\smallclaw(2,1,1)$ but it works in full generality.
For the three-star $\smallclaw(2,1,1)$ the graph $G$ has two edges (the edge attached to $z$ is contracted). Its graphical function is
$$
f_G(z)=\frac{1}{(z\zz)^{2\lambda}((z-1)(\zz-1))^\lambda}.
$$
The pole term in the expansion of the graphical function $f_G(z)$ on the right hand side of (\ref{eq:effLap}) at $z=0$ is given by (\ref{0asympt}).
The only term in the sum over $V$ is $V=\emptyset$. We get
$$
f_{G[V\cup\{0,z\}]}(z)=\frac{1}{(z\zz)^{2\lambda}}\qquad\text{and}\qquad f_{G[V\cup\{0,z\}=0]}=1.
$$
The general case has more terms in the sum over $V$ (we ignore all terms which do not have a pole of order $2n+4$ at $z=0$).
Before we append the edge we subtract the singular term(s). We hence apply the algorithm of Section \ref{sectappedge} to
$$
f_G(z)-f_{G[V\cup\{0,z\}]}(z)f_{G[V\cup\{0,z\}=0]}=\frac{1-((z-1)(\zz-1))^\lambda}{(z\zz)^{2\lambda}((z-1)(\zz-1))^\lambda}.
$$
The pole of order four at $|z|=0$ is lifted. The subtraction does not introduce new singularities at $z=1$ or at $z=\infty$ and we get the unique result
$$
f_{G_1}^{\mathrm{reg}}(z)=-\frac{(1+\epsilon(1-\frac\gamma2))\sL_1}{z\zz}-\epsilon\frac{(z-\zz)\sL_{11}+2z\sL_{01}-2\zz\sL_{10}}{2z\zz(z-\zz)}+\sO(\epsilon^2).
$$
We need to compensate for the subtraction by appending an edge to $f_{G[V\cup\{0,z\}]}(z)f_{G[V\cup\{0,z\}=0]}$. This singular term is the convolution
$$
f_{G_1}^{\mathrm{sing}}(z)=\frac{1}{\pi^{D/2}}\int_{\RR^D}\frac{\dd x}{|x|^{4\lambda}|x-z_2(z)|^{2\lambda}}.
$$
It can be calculated by Fourier transformation. We obtain
\begin{align}
f_{G_1}^{\mathrm{sing}}(z)&=\frac{1}{\Gamma(\lambda)(2\lambda-1)(1-\lambda)(z\zz)^{2\lambda-1}}\nonumber\\
&=\frac{2}{\epsilon z\zz}+\frac{2-\gamma+(2+\epsilon(2-\gamma))\sL_0}{z\zz}+\epsilon\frac{2-\gamma+\frac{\gamma^2}4-\frac{\pi^2}{24}+2\sL_{00}}{z\zz}+\sO(\epsilon^2).
\end{align}
The singular part of $f_{G_1}(z)$ contains a term $\sim (z\zz)^{-1}$ which is annihilated by $\Delta_0$ in (\ref{eq:effLap}).
The graphical function of $G_1$ is the sum of the regular and the singular part,
$$
f_{G_1}(z)=f_{G_1}^{\mathrm{reg}}(z)+f_{G_1}^{\mathrm{sing}}(z).
$$
In practice, we have to fix a maximum order $n$ in $\epsilon$ for $f_{G_1}(z)$. While the reduced graphical $f_G(z)$ needs only to be calculated to order $n$
we need to determine the subtraction terms $f_{G[V\cup\{0,z\}]}(z)f_{G[V\cup\{0,z\}=0]}$ to order $n+1$ because the singular convolution generates a pole in $\epsilon$.
Because the subtraction terms are mere two-point functions which are determined by Feynman periods (see Section \ref{sectper}) the extra order in $\epsilon$ typically can be handled.
Still, in particular in six-dimensional $\phi^3$ theory, the calculation of $f_{G_1}(z)$ can fail for a given $f_G(z)$ because the subtraction terms cannot be calculated.

The situation of a logarithmic pole at $z=1$ is fully analogous using the asymptotic formula
\begin{equation}\label{1asympt}
f_G(z)=\sum_{V\subseteq\sV^{\mathrm{int}}}f_{G[V\cup\{1,z\}]}(z)f_{G[V\cup\{1,z\}=1]}(1+O(|z-1|^2)).
\end{equation}

In case of a logarithmic singularity at infinity one has to subtract the asymptotic expansion \cite{numfunct}
\begin{equation}\label{inftyasympt}
f_G(z)=\sum_{V\subseteq\sV^{\mathrm{int}}}f_{G[V\cup\{0,1\}]}f_{G[V\cup\{0,1\}=0]}(z)(1+O(|z|^{-2})).
\end{equation}
As an example consider the three-star $\smallclaw(a\epsilon/\lambda,1,1)$ in Figure \ref{fig:3star} with an edge of weight $a\epsilon/\lambda$ attached to $0$ (for some $a\in\RR$).
We obtain for the graph $G$ the function
$$
f_G(z)=\frac{1}{(z\zz)^{a\epsilon}((z-1)(\zz-1))^\lambda},
$$
which is of order $|z|^{-2}$ for $z\to\infty$ (and $\epsilon=0$). The expansion of $f_G(z)$ at $z=\infty$ is given by (\ref{inftyasympt}) with $V=\emptyset$. We get
$$
f_{G[V\cup\{0,1\}]}=1\qquad\text{and}\qquad f_{G[V\cup\{0,1\}=0]}(z)=\frac{1}{(z\zz)^{\lambda+a\epsilon}}.
$$
Subtraction of the logarithmic pole at infinity gives
$$
f_G(z)-f_{G[V\cup\{0,1\}]}f_{G[V\cup\{0,1\}=0]}(z)=\frac{1-((1-1/z)(1-1/\zz))^\lambda}{(z\zz)^{a\epsilon}((z-1)(\zz-1))^\lambda}.
$$
After the subtraction the function is of order $|z|^{-4}$ for $z\to\infty$. Moreover, the subtraction does not induce poles of order $\geq4$ at $|z|=0$ (or at $z=1$).
Integration with the algorithm of Section \ref{sectappedge} gives the unique result
$$
f_{G_1}^{\mathrm{reg}}(z)=\Big(1+\epsilon\Big(a-\frac\gamma2\Big)\Big)(\sL_0-\sL_1)+\epsilon\Big(\frac{(1-2a)\sL_{00}-\sL_{11}}{2}+\frac{-a\zz\sL_{01}+az\sL_{10}}{z-\zz}\Big)+\sO(\epsilon^2).
$$
The contribution from the subtraction is a convolution, yielding
\begin{equation}\label{fGs}
f_{G_1}^{\mathrm{sing}}(z)=-\Big(1+\epsilon\Big(a-\frac\gamma2\Big)\Big)\sL_0-\epsilon\frac{(1-2a)\sL_{00}}{2}+\frac{-\frac2\epsilon-2a+\gamma+\epsilon(-2a^2+a\gamma-\frac{\gamma^2}4+\frac{\pi^2}{24})}{1-2a}
+\sO(\epsilon^2).
\end{equation}
The constant term in $f_{G_1}^{\mathrm{sing}}(z)$ is annihilated by $\Delta_0$ in (\ref{eq:effLap}).
The function of $G_1$ is the sum of the regular and the singular part, $f_{G_1}(z)=f_{G_1}^{\mathrm{reg}}(z)+f_{G_1}^{\mathrm{sing}}(z)$.

This example also illustrates that it is possible to calculate graphical functions with weights where the coefficient of $\epsilon$ is a free parameter.
This gives families of graphical functions with fixed weights in the limit $\epsilon=0$. Knowledge of such families is quite helpful for the calculation
of renormalization functions at high loop orders.

Note that (\ref{fGs}) is singular for $a=1/2$. The graphical function $f_{G_1}(z)$ does not exist for this value of $a$. Although $\epsilon$ is a regulator, not all Feynman
integrals exist in $2n+4-\epsilon$ dimensions.

The most general weight of an edge $e$ that one needs for the calculation of renormalization functions is
\begin{equation}\label{nua}
\nu(n_e,a_e)=\frac{n_e+a_e\epsilon}{\lambda},\quad\text{with }n_e\in\ZZ\quad\text{and}\quad a_e\in\QQ.
\end{equation}

\section{Two-point graphs and periods}\label{sectper}

In the previous sections we explained the fundamental algorithms for the computation of $\epsilon$-expansions of graphical functions. In practical situations, however,
one often is merely interested in the calculation of a dimensionally regularized two-point function $A_G(z_1,z_2)$. By (\ref{AGinvs}) and (\ref{NGdef}) a two-point integral is determined
by its scaling behavior and a number (for fixed $\epsilon$) which we call `the period of $G$',
\begin{equation}\label{PGinvs}
A_G(z_0,z_1)=||z_1-z_0||^{-2\lambda N_G}P_G.
\end{equation}
Note that $P_G=f_G(z)$ is a constant graphical function because the absence of the vertex $z_2$ eliminates the dependence on $z$ in (\ref{eqzdef}).
When we consider $G$ as the graph of a constant graphical function, we add an isolated vertex $z$ to $G$ (which also has external vertices $0$ and $1$).

\begin{figure}
\centering
\includegraphics{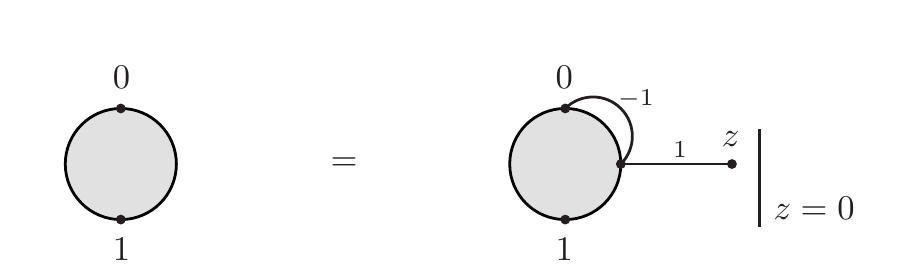}
\caption{Integration over $z$ by appending an edge of weight $1$.}
\label{fig:int}
\end{figure}

In general, periods are more accessible than graphical functions. To calculate $P_G$ we may choose any internal vertex in the graph $G$ and make it external by giving it the
label $z$ (rather than adding an isolated vertex $z$). If it is possible to calculate any such graphical function, we can integrate over $z$ to obtain the period $P_G$.
In integer dimensions this integration over $z$ is best facilitated by using a residue theorem (Theorem 2.29 in \cite{gf} with generalization to even dimensions $\geq4$ in \cite{gfe}).
The residue theorem is very efficient, but it does not generalize straightforwardly to non-integer dimensions: The integrand fails to be a GSVH (because it is not single-valued)
even if the graphical function is. On the other hand, in non-integer dimensions one is typically interested in much lower loop orders. This enables us to use a simple work-around
for the integration over $z$. We add an edge from $z$ to $0$ of weight $-1$ and append an edge of weight $1$ to $z$, see Figure \ref{fig:int}. Then, we set $z=0$ in the resulting
graphical function. This cancels the factor $(z\zz)^\lambda$ from the edge $z0$ of weight $-1$ and we obtain $P_G$. The algorithm for appending an edge typically is efficient
(see Section \ref{sectappedge}). So, the work-around is acceptable although quite some extra information is calculated in the appended graphical function before we set $z=0$ \cite{numfunct}.

\section{Constructible graphs}\label{sectcon}

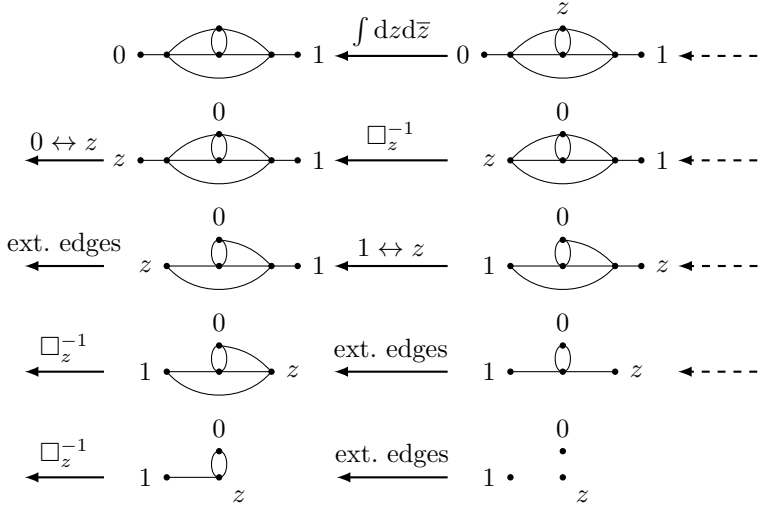
\begin{figure}
\centering
\begin{tikzpicture}[x=2.3ex,y=2.3ex]

    \begin{scope}[local bounding box=fullcateye]
        \coordinate (vm);
        \coordinate [above=1 of vm] (vo);
        \coordinate [left =2 of vm] (vl);
        \coordinate [right=2 of vm] (vr);
        \coordinate [left =1 of vl] (vll);
        \coordinate [right=1 of vr] (vrr);

        \draw (vm) to [bend left =90] (vo);
        \draw (vm) to [bend right=90] (vo);
        \draw (vl) to [bend left =20] (vo);
        \draw (vr) to [bend right=20] (vo);
        \draw (vl) to [bend right=50] (vr);

        \draw (vl) -- (vm);
        \draw (vr) -- (vm);
        \draw (vll) -- (vl);
        \draw (vrr) -- (vr);

        \node [left =.2 of vll] {$0$};
        \node [right=.2 of vrr] {$1$};

        \filldraw (vm) circle (1pt);
        \filldraw (vo) circle (1pt);
        \filldraw (vl) circle (1pt);
        \filldraw (vr) circle (1pt);
        \filldraw (vll) circle (1pt);
        \filldraw (vrr) circle (1pt);
    \end{scope}

    \begin{scope}[xshift=130,local bounding box=fullcateyez]
        \coordinate (vm);
        \coordinate [above=1 of vm] (vo);
        \coordinate [left =2 of vm] (vl);
        \coordinate [right=2 of vm] (vr);
        \coordinate [left =1 of vl] (vll);
        \coordinate [right=1 of vr] (vrr);

        \draw (vm) to [bend left =90] (vo);
        \draw (vm) to [bend right=90] (vo);
        \draw (vl) to [bend left =20] (vo);
        \draw (vr) to [bend right=20] (vo);
        \draw (vl) to [bend right=50] (vr);

        \draw (vl) -- (vm);
        \draw (vr) -- (vm);
        \draw (vll) -- (vl);
        \draw (vrr) -- (vr);

        \node [left =.2 of vll] {$0$};
        \node [right=.2 of vrr] {$1$};
        \node [above=.2 of vo ] {$z$};

        \filldraw (vm) circle (1pt);
        \filldraw (vo) circle (1pt);
        \filldraw (vl) circle (1pt);
        \filldraw (vr) circle (1pt);
        \filldraw (vll) circle (1pt);
        \filldraw (vrr) circle (1pt);
    \end{scope}

    \draw[-latex,thick] (fullcateye-|fullcateyez.west) -- (fullcateye) node[midway,above]{$\int \dd z \dd \zz$};
    \draw[-latex,thick,dashed]  ([xshift=30]fullcateye-|fullcateyez.east) -- (fullcateye-|fullcateyez.east)  node[midway,above]{};

    \begin{scope}[yshift=-40,local bounding box=fullcateyez2]
        \coordinate (vm);
        \coordinate [above=1 of vm] (vo);
        \coordinate [below=1 of vm] (vu);
        \coordinate [left =2 of vm] (vl);
        \coordinate [right=2 of vm] (vr);
        \coordinate [left =1 of vl] (vll);
        \coordinate [right=1 of vr] (vrr);

        \draw (vm) to [bend left =90] (vo);
        \draw (vm) to [bend right=90] (vo);
        \draw (vl) to [bend left =20] (vo);
        \draw (vr) to [bend right=20] (vo);
        \draw (vl) to [bend right=50] (vr);

        \draw (vl) -- (vm);
        \draw (vr) -- (vm);
        \draw (vll) -- (vl);
        \draw (vrr) -- (vr);

        \node [left =.2 of vll] {$z$};
        \node [right=.2 of vrr] {$1$};
        \node [above=.2 of vo ] {$0$};

        \node [below=.2 of vu ] {\phantom{$0$}};

        \filldraw (vm) circle (1pt);
        \filldraw (vo) circle (1pt);
        \filldraw (vl) circle (1pt);
        \filldraw (vr) circle (1pt);
        \filldraw (vll) circle (1pt);
        \filldraw (vrr) circle (1pt);
    \end{scope}

    \draw[-latex,thick]  (fullcateyez2) -- ([xshift=-30]fullcateyez2-|fullcateyez2.west)  node[midway,above]{$0 \leftrightarrow z$};


    \begin{scope}[yshift=-40,xshift=130,local bounding box=cateyeredl]
        \coordinate (vm);
        \coordinate [above=1 of vm] (vo);
        \coordinate [below=1 of vm] (vu);
        \coordinate [left =2 of vm] (vl);
        \coordinate [right=2 of vm] (vr);
        \coordinate [left =1 of vl] (vll);
        \coordinate [right=1 of vr] (vrr);

        \draw (vm) to [bend left =90] (vo);
        \draw (vm) to [bend right=90] (vo);
        \draw (vl) to [bend left =20] (vo);
        \draw (vr) to [bend right=20] (vo);
        \draw (vl) to [bend right=50] (vr);

        \draw (vl) -- (vm);
        \draw (vr) -- (vm);
        \draw (vrr) -- (vr);

        \node [left =.2 of vl] {$z$};
        \node [right=.2 of vrr] {$1$};
        \node [above=.2 of vo ] {$0$};
        \node [below=.2 of vu ] {\phantom{$0$}};
        \node [right=.2 of vrr ] {\phantom{$0$}};
        \node [left=.2 of vll ] {\phantom{$0$}};

        \filldraw (vm) circle (1pt);
        \filldraw (vo) circle (1pt);
        \filldraw (vl) circle (1pt);
        \filldraw (vr) circle (1pt);
        \filldraw (vrr) circle (1pt);
    \end{scope}

    \draw[-latex,thick] (fullcateyez2-|cateyeredl.west) -- (fullcateyez2) node[midway,above]{$\Box_{z}^{-1}$};
    \draw[-latex,thick,dashed]  ([xshift=30]fullcateyez2-|cateyeredl.east) -- (fullcateyez2-|cateyeredl.east)  node[midway,above]{};

    \begin{scope}[yshift=-80,xshift=0,local bounding box=cateyeredl2]
        \coordinate (vm);
        \coordinate [above=1 of vm] (vo);
        \coordinate [below=1 of vm] (vu);
        \coordinate [left =2 of vm] (vl);
        \coordinate [right=2 of vm] (vr);
        \coordinate [left =1 of vl] (vll);
        \coordinate [right=1 of vr] (vrr);

        \draw (vm) to [bend left =90] (vo);
        \draw (vm) to [bend right=90] (vo);
        \draw (vr) to [bend right=20] (vo);
        \draw (vl) to [bend right=50] (vr);

        \draw (vl) -- (vm);
        \draw (vr) -- (vm);
        \draw (vrr) -- (vr);

        \node [left =.2 of vl] {$z$};
        \node [right=.2 of vrr] {$1$};
        \node [above=.2 of vo ] {$0$};
        \node [below=.2 of vu ] {\phantom{$0$}};
        \node [right=.2 of vrr ] {\phantom{$0$}};
        \node [left=.2 of vll ] {\phantom{$0$}};

        \filldraw (vm) circle (1pt);
        \filldraw (vo) circle (1pt);
        \filldraw (vl) circle (1pt);
        \filldraw (vr) circle (1pt);
        \filldraw (vrr) circle (1pt);
    \end{scope}

    \draw[-latex,thick]  (cateyeredl2) -- ([xshift=-30]cateyeredl2-|cateyeredl2.west)  node[midway,above]{\text{ext.~edges}};

    \begin{scope}[yshift=-80,xshift=130,local bounding box=cateyeredl3]
        \coordinate (vm);
        \coordinate [above=1 of vm] (vo);
        \coordinate [below=1 of vm] (vu);
        \coordinate [left =2 of vm] (vl);
        \coordinate [right=2 of vm] (vr);
        \coordinate [left =1 of vl] (vll);
        \coordinate [right=1 of vr] (vrr);

        \draw (vm) to [bend left =90] (vo);
        \draw (vm) to [bend right=90] (vo);
        \draw (vr) to [bend right=20] (vo);
        \draw (vl) to [bend right=50] (vr);

        \draw (vl) -- (vm);
        \draw (vr) -- (vm);
        \draw (vrr) -- (vr);

        \node [left =.2 of vl] {$1$};
        \node [right=.2 of vrr] {$z$};
        \node [right=.2 of vrr ] {\phantom{$0$}};
        \node [left=.2 of vll ] {\phantom{$0$}};
        \node [above=.2 of vo ] {$0$};
        \node [below=.2 of vu ] {\phantom{$0$}};

        \filldraw (vm) circle (1pt);
        \filldraw (vo) circle (1pt);
        \filldraw (vl) circle (1pt);
        \filldraw (vr) circle (1pt);
        \filldraw (vrr) circle (1pt);
    \end{scope}

    \draw[-latex,thick]  (cateyeredl2-|cateyeredl3.west) -- (cateyeredl2)  node[midway,above]{$1 \leftrightarrow z$};

    \draw[-latex,thick,dashed]  ([xshift=30]cateyeredl3-|cateyeredl3.east) -- (cateyeredl3)  node[midway,above]{};

    \begin{scope}[yshift=-120,local bounding box=cateyeredlr]
        \coordinate (vm);
        \coordinate [above=1 of vm] (vo);
        \coordinate [below=1 of vm] (vu);
        \coordinate [left =2 of vm] (vl);
        \coordinate [right=2 of vm] (vr);
        \coordinate [left =1 of vl] (vll);
        \coordinate [right=1 of vr] (vrr);

        \draw (vm) to [bend left =90] (vo);
        \draw (vm) to [bend right=90] (vo);
        \draw (vr) to [bend right=20] (vo);
        \draw (vl) to [bend right=50] (vr);

        \draw (vl) -- (vm);
        \draw (vr) -- (vm);

        \node [left =.2 of vl] {$1$};
        \node [right=.2 of vr] {$z$};
        \node [above=.2 of vo ] {$0$};
        \node [below=.2 of vu ] {\phantom{$0$}};
        \node [right=.2 of vrr ] {\phantom{$0$}};
        \node [left=.2 of vll ] {\phantom{$0$}};

        \filldraw (vm) circle (1pt);
        \filldraw (vo) circle (1pt);
        \filldraw (vl) circle (1pt);
        \filldraw (vr) circle (1pt);
    \end{scope}

    \draw[-latex,thick]  (cateyeredlr) -- ([xshift=-30]cateyeredlr-|cateyeredlr.west)  node[midway,above]{$\Box_{z}^{-1}$};

    \begin{scope}[yshift=-120,xshift=130,local bounding box=cateyeredlr2]
        \coordinate (vm);
        \coordinate [above=1 of vm] (vo);
        \coordinate [below=1 of vm] (vu);
        \coordinate [left =2 of vm] (vl);
        \coordinate [right=2 of vm] (vr);
        \coordinate [left =1 of vl] (vll);
        \coordinate [right=1 of vr] (vrr);

        \draw (vm) to [bend left =90] (vo);
        \draw (vm) to [bend right=90] (vo);

        \draw (vl) -- (vm);
        \draw (vr) -- (vm);

        \node [left =.2 of vl] {$1$};
        \node [right=.2 of vr] {$z$};
        \node [above=.2 of vo ] {$0$};
        \node [below=.2 of vu ] {\phantom{$0$}};
        \node [right=.2 of vrr ] {\phantom{$0$}};
        \node [left=.2 of vll ] {\phantom{$0$}};

        \filldraw (vm) circle (1pt);
        \filldraw (vo) circle (1pt);
        \filldraw (vl) circle (1pt);
        \filldraw (vr) circle (1pt);
    \end{scope}

    \draw[-latex,thick]  (cateyeredlr-|cateyeredlr2.west) -- (cateyeredlr)  node[midway,above]{\text{ext.~edges}};
    \draw[-latex,thick,dashed]  ([xshift=30]cateyeredlr2-|cateyeredlr2.east) -- (cateyeredlr2)  node[midway,above]{};

    \begin{scope}[yshift=-160,xshift=0,local bounding box=cateyeredlr3]
        \coordinate (vm);
        \coordinate [above=1 of vm] (vo);
        \coordinate [below=1 of vm] (vu);
        \coordinate [left =2 of vm] (vl);
        \coordinate [right=2 of vm] (vr);
        \coordinate [left =1 of vl] (vll);
        \coordinate [right=1 of vr] (vrr);

        \draw (vm) to [bend left =90] (vo);
        \draw (vm) to [bend right=90] (vo);

        \draw (vl) -- (vm);

        \node [left =.2 of vl] {$1$};
        \node [below right=.2 of vm] {$z$};
        \node [above=.2 of vo ] {$0$};
        \node [below=.2 of vu ] {\phantom{$0$}};
        \node [right=.2 of vrr ] {\phantom{$0$}};
        \node [left=.2 of vll ] {\phantom{$0$}};

        \filldraw (vm) circle (1pt);
        \filldraw (vo) circle (1pt);
        \filldraw (vl) circle (1pt);
    \end{scope}

    \draw[-latex,thick]  (cateyeredlr3) -- ([xshift=-30]cateyeredlr3-|cateyeredlr3.west)  node[midway,above]{$\Box_{z}^{-1}$};


    \begin{scope}[yshift=-160,xshift=130,local bounding box=one]
        \coordinate (vm);
        \coordinate [above=1 of vm] (vo);
        \coordinate [below=1 of vm] (vu);
        \coordinate [left =2 of vm] (vl);
        \coordinate [right=2 of vm] (vr);
        \coordinate [left =1 of vl] (vll);
        \coordinate [right=1 of vr] (vrr);



        \node [left =.2 of vl] {$1$};
        \node [below right=.2 of vm] {$z$};
        \node [above=.2 of vo ] {$0$};
        \node [below=.2 of vu ] {\phantom{$0$}};
        \node [right=.2 of vrr ] {\phantom{$0$}};
        \node [left=.2 of vll ] {\phantom{$0$}};

        \filldraw (vm) circle (1pt);
        \filldraw (vo) circle (1pt);
        \filldraw (vl) circle (1pt);
    \end{scope}

    \draw[-latex,thick]  (one) -- ([xshift=-42]one-|one.west)  node[midway,above]{\text{ext.~edges}};
\end{tikzpicture}

\caption{Example of the graphical reduction of a two-point function in $\phi^4$ theory.
The explicit expression can be obtained by following the arrows which start from the trivial graphical function which is equal to $1$. The symbol $\Box_{z}^{-1}$ means appending an edge
and $\int\dd z\dd\zz$ is integration over $z$.}
\label{fig:GFreduction}
\end{figure}
The core integration technique for constructing graphical functions is appending edges, see Section \ref{sectappedge}. The technique becomes quite powerful in combination
with four elementary identities.

Because in graphical functions we do not integrate over variables associated to external vertices, adding an external edge of weight $\nu$ merely gives a propagator factor, see (\ref{propagator}). Explicitly,
\begin{align*}
\left[
\begin{tikzpicture}[x=1.5ex,y=1.5ex,baseline={([yshift=-.6ex]current bounding box.center)}]
    \coordinate (v0);
    \coordinate [above right=1.41 and 2 of v0] (v1);
    \coordinate [below right=1.41 and 2 of v0] (v2);
    \node [left =.1 of v0] (v0n) { $z$};
    \coordinate [right =1 of v0] (v0u);
    \node [right=.07 of v1] (v1n) { $1$};
    \coordinate [below left=.71 and .71 of v1] (v1u);
    \node [right=.07 of v2] (v2n) { $0$};
    \coordinate [above left=.71 and .71 of v2] (v2u);
    \draw[fill=gray] (v0) .. controls (v0u) and (v1u) .. (v1) .. controls (v1u) and (v2u) .. (v2) .. controls (v2u) and (v0u) .. (v0);
    \filldraw (v0) circle (1pt);
    \filldraw (v1) circle (1pt);
    \filldraw (v2) circle (1pt);
\end{tikzpicture}%
\right]
&=
\left[
\begin{tikzpicture}[x=1.5ex,y=1.5ex,baseline={([yshift=-.6ex]current bounding box.center)}]
    \coordinate (v0);
    \coordinate [above right=1.41 and 2 of v0] (v1);
    \coordinate [below right=1.41 and 2 of v0] (v2);
    \node [left =.1 of v0] (v0n) {$z$};
    \coordinate [right =1 of v0] (v0u);
    \node [right=.07 of v1] (v1n) { $1$};
    \coordinate [below left=.71 and .71 of v1] (v1u);
    \node [right=.07 of v2] (v2n) { $0$};
    \coordinate [above left=.71 and .71 of v2] (v2u);
    \draw[fill=gray] (v0) .. controls (v0u) and (v1u) .. (v1) .. controls (v1u) and (v2u) .. (v2) .. controls (v2u) and (v0u) .. (v0);
    \draw (v1) to [bend left=90] (v2);
    \filldraw (v0) circle (1pt);
    \filldraw (v1) circle (1pt);
    \filldraw (v2) circle (1pt);
\end{tikzpicture}%
\right]
=
(z\zz)^{\lambda\nu}
\left[
\begin{tikzpicture}[x=1.5ex,y=1.5ex,baseline={([yshift=-.6ex]current bounding box.center)}]
    \coordinate (v0);
    \coordinate [above right=1.41 and 2 of v0] (v1);
    \coordinate [below right=1.41 and 2 of v0] (v2);
    \node [left =.1 of v0] (v0n) { $z$};
    \coordinate [right =1 of v0] (v0u);
    \node [right=.07 of v1] (v1n) { $1$};
    \coordinate [below left=.71 and .71 of v1] (v1u);
    \node [right=.07 of v2] (v2n) { $0$};
    \coordinate [above left=.71 and .71 of v2] (v2u);
    \draw[fill=gray] (v0) .. controls (v0u) and (v1u) .. (v1) .. controls (v1u) and (v2u) .. (v2) .. controls (v2u) and (v0u) .. (v0);
    \draw (v2) to [bend left=90] (v0);
    \filldraw (v0) circle (1pt);
    \filldraw (v1) circle (1pt);
    \filldraw (v2) circle (1pt);
\end{tikzpicture}%
\right]
=
\left[(z-1)(\zz-1) \right]^{\lambda\nu}
\left[
\begin{tikzpicture}[x=1.5ex,y=1.5ex,baseline={([yshift=-.6ex]current bounding box.center)}]
    \coordinate (v0);
    \coordinate [above right=1.41 and 2 of v0] (v1);
    \coordinate [below right=1.41 and 2 of v0] (v2);
    \node [left =.1 of v0] (v0n) { $z$};
    \coordinate [right =1 of v0] (v0u);
    \node [right=.07 of v1] (v1n) { $1$};
    \coordinate [below left=.71 and .71 of v1] (v1u);
    \node [right=.07 of v2] (v2n) { $0$};
    \coordinate [above left=.71 and .71 of v2] (v2u);
    \draw[fill=gray] (v0) .. controls (v0u) and (v1u) .. (v1) .. controls (v1u) and (v2u) .. (v2) .. controls (v2u) and (v0u) .. (v0);
    \draw (v0) to [bend left=90] (v1);
    \filldraw (v0) circle (1pt);
    \filldraw (v1) circle (1pt);
    \filldraw (v2) circle (1pt);
\end{tikzpicture}%
\right].
\end{align*}
A permutation of the three external vertices $0$, $1$, $z$ results in a M\"obius transformation of the argument and a scaling factor,
\begin{align}\label{trafos}
\left[
\begin{tikzpicture}[x=1.5ex,y=1.5ex,baseline={([yshift=-.6ex]current bounding box.center)}]
    \coordinate (v0);
    \coordinate [above right=1.41 and 2 of v0] (v1);
    \coordinate [below right=1.41 and 2 of v0] (v2);
    \node [left =.1 of v0] (v0n) { $z$};
    \coordinate [right =1 of v0] (v0u);
    \node [right=.07 of v1] (v1n) { $0$};
    \coordinate [below left=.71 and .71 of v1] (v1u);
    \node [right=.07 of v2] (v2n) { $1$};
    \coordinate [above left=.71 and .71 of v2] (v2u);
    \draw[fill=gray] (v0) .. controls (v0u) and (v1u) .. (v1) .. controls (v1u) and (v2u) .. (v2) .. controls (v2u) and (v0u) .. (v0);
    \filldraw (v0) circle (1pt);
    \filldraw (v1) circle (1pt);
    \filldraw (v2) circle (1pt);
\end{tikzpicture}%
\right]
&=
\left[
\begin{tikzpicture}[x=1.5ex,y=1.5ex,baseline={([yshift=-.6ex]current bounding box.center)}]
    \coordinate (v0);
    \coordinate [above right=1.41 and 2 of v0] (v1);
    \coordinate [below right=1.41 and 2 of v0] (v2);
    \node [left =.1 of v0] (v0n) { $1\!-\!z$};
    \coordinate [right =1 of v0] (v0u);
    \node [right=.07 of v1] (v1n) { $1$};
    \coordinate [below left=.71 and .71 of v1] (v1u);
    \node [right=.07 of v2] (v2n) { $0$};
    \coordinate [above left=.71 and .71 of v2] (v2u);
    \draw[fill=gray] (v0) .. controls (v0u) and (v1u) .. (v1) .. controls (v1u) and (v2u) .. (v2) .. controls (v2u) and (v0u) .. (v0);
    \filldraw (v0) circle (1pt);
    \filldraw (v1) circle (1pt);
    \filldraw (v2) circle (1pt);
\end{tikzpicture}%
\right]
=
(z\zz)^{-\lambda N_G}
\left[
\begin{tikzpicture}[x=1.5ex,y=1.5ex,baseline={([yshift=-.6ex]current bounding box.center)}]
    \coordinate (v0);
    \coordinate [above right=1.41 and 2 of v0] (v1);
    \coordinate [below right=1.41 and 2 of v0] (v2);
    \node [left =.1 of v0] (v0n) { $1$};
    \coordinate [right =1 of v0] (v0u);
    \node [right=.07 of v1] (v1n) { $0$};
    \coordinate [below left=.71 and .71 of v1] (v1u);
    \node [right=.07 of v2] (v2n) { $\frac{1}{z}$};
    \coordinate [above left=.71 and .71 of v2] (v2u);
    \draw[fill=gray] (v0) .. controls (v0u) and (v1u) .. (v1) .. controls (v1u) and (v2u) .. (v2) .. controls (v2u) and (v0u) .. (v0);
    \filldraw (v0) circle (1pt);
    \filldraw (v1) circle (1pt);
    \filldraw (v2) circle (1pt);
\end{tikzpicture}%
\right].
\end{align}
This identity is an immediate consequence of permuting the interpretation of $z_0$, $z_1$, $z_2$ in the Feynman integral $A_G(z_0,z_1,z_2)$
and using the representation (\ref{AGinvs}) of a graphical function in terms of the invariants (\ref{eqinvs}).
The two transformations $(0,1,z)\leftrightarrow(1,0,1-z)$ and $(0,1,z)\leftrightarrow(0,1/z,1)$ generate the six elements of the permutation group that permutes $0$, $1$, and $z$.
The transformation of GSVHs under $z\to1-z$ and $z\to1/z$ can be calculated \cite{GSVH}.

The third identity applies in the rather rare case that the graph $G$ consists of two subgraphs $G_1$ and $G_2$ that only connect at the external vertices $0$, $1$, and $z$. In this case
$$
f_G(z)=f_{G_1}(z)f_{G_2}(z)
$$
is the product of the graphical functions of $G_1$ and $G_2$. This product identity holds because the integrand of $A_G(z_0,z_1,z_2)$ factorizes into disjoint sets of variables according to the internal vertices of $G_1$ and $G_2$.

Finally, by scaling, a two-point insertion $g$ factorizes into an edge with weight $N_g$, see (\ref{NGdef}), times a Feynman period, see (\ref{PGinvs}).
If the Feynman period can be calculated with the method in Section \ref{sectper} using constructible graphical functions, then the two-point insertion is constructible
and it can be used to calculate constructible graphical functions. A simple example is the chain of two weighted edges where we get
\begin{align}\label{eqconvolute}
&\begin{tikzpicture}[baseline={([yshift=-2.5ex]current bounding box.center)}]
    \coordinate (v1) at (0,0);
    \coordinate (v2) at (1.5,0);
    \coordinate (v3) at (3,0);
    \filldraw[black!20] (v1) -- ([shift=(150:.2)]v1) arc (150:210:.2) -- (v1);
    \draw (v1) -- ([shift=(150:.2)]v1);
    \draw (v1) -- ([shift=(210:.2)]v1);
    \filldraw[black!20] (v3) -- ([shift=(30:.2)]v3) arc (30:-30:.2) -- (v3);
    \draw (v3) -- ([shift=(30:.2)]v3);
    \draw (v3) -- ([shift=(-30:.2)]v3);
    \filldraw (v1) circle (1.3pt);
    \filldraw (v2) circle (1.3pt);
    \filldraw (v3) circle (1.3pt);
    \draw[white] (v1) -- node[above] {$\phantom{\nu_1+\nu_2-\frac{\lambda+1}{\lambda}}$} (v3);
    \draw (v1) -- node[above] {$\nu_1$} (v2);
    \draw (v2) -- node[above] {$\nu_2$} (v3);
\end{tikzpicture}
\quad
=
\quad
c^{(\lambda)}_{\nu_1,\nu_2}
\quad
\begin{tikzpicture}[baseline={([yshift=-2.5ex]current bounding box.center)}]
    \coordinate (v1) at (0,0);
    \coordinate (v3) at (3,0);
    \filldraw[black!20] (v1) -- ([shift=(150:.2)]v1) arc (150:210:.2) -- (v1);
    \draw (v1) -- ([shift=(150:.2)]v1);
    \draw (v1) -- ([shift=(210:.2)]v1);
    \filldraw[black!20] (v3) -- ([shift=(30:.2)]v3) arc (30:-30:.2) -- (v3);
    \draw (v3) -- ([shift=(30:.2)]v3);
    \draw (v3) -- ([shift=(-30:.2)]v3);
    \filldraw (v1) circle (1.3pt);
    \filldraw (v3) circle (1.3pt);
    \draw (v1) -- node[above] {$\nu_1+\nu_2-\frac{\lambda+1}{\lambda}$} (v3);
\end{tikzpicture}\\
&\text{where } \quad
c^{(\lambda)}_{\nu_1,\nu_2} = \frac{\Gamma(\lambda(1-\nu_1)+1)\Gamma(\lambda(1-\nu_2)+1)\Gamma(\lambda(\nu_1+\nu_2-1)-1)}
{\Gamma(\lambda\nu_1)\Gamma(\lambda\nu_2)\Gamma(\lambda(2-\nu_1-\nu_2)+2)},\nonumber
\end{align}
whenever $c^{(\lambda)}_{\nu_1,\nu_2}$ exists.

We call a graph $G$ {\em constructible} if the reductions
\begin{enumerate}
\item deletion of external edges,
\item permutation of the three external vertices $0$, $1$, $z$,
\item product factorization,
\item factorization of constructible two-point insertions,
\item deletion of a single weight 1 edge at the vertex $z$
\end{enumerate}
reduces $G$ to empty graphical functions (which consist of the vertices $0$, $1$, $z$ with no edges). The empty graphical function has the constant value 1.

By inverting the reduction steps, constructible graphical functions in $2n+4-\epsilon$ dimensions can in principle be calculated to any order in $\epsilon$.
In practice, one encounters time and memory limitations. These limitation, however, are mild in the sense that they do not limit the calculation of constructible graphical functions
at loop orders which are relevant for the calculation of renormalization functions.

A period $P_G$ is constructible if there exists an internal vertex $x$ in $G$ such that the graphical function of the graph $G|_{x=z}$ (where $x$ is replaced by the external vertex
$z$) is contructible. This definition is justified because the algorithm for the integration over $z$ (see Section \ref{sectper}) only adds an external edge
and appends an edge of weight 1 to $z$. The final limit $z=0$ can easily be calculated.

Any $\epsilon$-coefficient of any constructible graphical function is a GSVH in the alphabet $0$, $1$, $\zz$ \cite{GSVH}.
Any $\epsilon$-coefficient of any constructible period is a multiple zeta value (MZV).

Many graphical functions and periods are constructible. A detailed example for the construction of the cat eye graph is in Figure \ref{fig:GFreduction}.

In \cite{gf} constructible graphs in four dimensions were introduced as graphs whose functions can be calculated by a larger set of reduction steps. The concept was
refined and generalized to $2n+4$ dimensions in \cite{gfe}. The reductions are (Section 7 in \cite{gfe})
\begin{enumerate}
\item completion with external vertex $\infty$,
\item deletion of external edges,
\item permutation of the four external vertices $0$, $1$, $z$, $\infty$,
\item product factorization,
\item factorization of constructible periods,
\item deletion of a single edge of weight $\in\ZZ/\lambda$ at the vertex $z$.
\end{enumerate}
In integer dimensions a graph is constructible if and only if it reduces with these steps to empty graphical functions.
Again, the idea is to define a class of graphical functions that can be calculated (at high loop orders subject to memory and time constraints).
The larger set of reduction steps uses conformal symmetry in integer dimensions (see Section \ref{sectid} for completion).

In $2n-4-\epsilon$ dimensions the extra reductions cannot be used directly.
Only after the application of other techniques (see Section \ref{sectapp}) some of these reductions become valuable. In Sections \ref{sectid} and \ref{sectapp} we demonstrate
how completion in combination with permutation of the four external vertices can be used to calculate the first orders in the $\epsilon$-expansion of some graphical functions.

\section{Identities}\label{sectid}
Graphical functions fulfill a plethora of identities in integer dimensions $\geq4$ that generalize to $2n+4-\epsilon$ dimensions \cite{gfe}.
Many of these identities are neither studied in detail nor implemented in the Maple package {\tt HyperlogProcedures} \cite{Shlog}.

The most important identities in non-integer dimensions are completion and integration by parts.
\subsection{Completion}\label{sectcom}
Completion is a consequence of conformal invariance in integer dimensions \cite{gf,gfe}. An external vertex `$\infty$' is added to the graph $G$ with edges such that all internal
vertices have the weight $D/\lambda$ (i.e.\ they have conformal weight).
For graphical functions one also adds edges $z\infty$, $01$, $0\infty$, $1\infty$ whose weights are adjusted such that all external vertices
have weight $0$. In completed Feynman periods all vertices have weight $D/\lambda$ \cite{numfunct}. After completion, the permutation symmetry of external vertices include the vertex $\infty$.
For graphical functions one obtains a 24-fold symmetry group of permutations of $0,1,z,\infty$ where the vertex $z$ undergoes a M\"obius transformation $\phi(z)$ such that the cross ratio
$$
\frac{(z_3-z_1)(z_2-z_0)}{(z_3-z_0)(z_2-z_1)}
$$
is stabilized with $\{z_0,z_1,z_2,z_3\}=\{0,1,\phi(z),\infty\}$ (in any order). Note that in double transpositions (like $0\leftrightarrow1$, $z\leftrightarrow\infty$) the variable
$z$ is not transformed.

In integer dimensions, completion is a very powerful tool to relate graphical functions of different graphs (when $\infty$ is moved from one vertex to another).
In particular, if many internal vertices have the weight $D/\lambda$, then the vertex $\infty$ connects only to the few non-conformal internal vertices.
If $\infty$ in the completed graph connects to no internal vertices, the graphical function becomes a Feynman period after $0\leftrightarrow1$, $z\leftrightarrow\infty$.
If $\infty$ connects to a single internal vertex, then the graph has an appended vertex $z$ after $0\leftrightarrow1$, $z\leftrightarrow\infty$.
Its graphical function can be calculated from the reduced graph (in integer dimensions appending an edge is also possible if the appended edge has weight $\neq1$ \cite{gfe}).
Otherwise, it may be possible to reduce the number of edges of a graphical function by moving $\infty$ to an external vertex with high valence, see Section \ref{sectex}.

In non-integer dimensions, vertices with weight 1 edges are never conformal (the regularization breaks conformal symmetry).
In four-dimensional $\phi^4$ theory ($n=0$) or in six-dimensional $\phi^3$ theory ($n=1$) any internal vertex $x$ has weight $(2n+4)/(n+1)$. This implies that $x$
connects to $\infty$ with the weight
\begin{equation}\label{wtinf}
\nu_{x\infty}=\frac{D}{\lambda}-\frac{2n+4}{n+1}=\frac{\epsilon}{(n+1)\lambda}.
\end{equation}
A priori, no simplifications are expected from completion. It is, however, possible to utilize the fact that the weight of edges
from internal vertices to $\infty$ have weights of low degree $\epsilon$ if one is only interested in low order expansions.

As an example consider the three-star in four dimensions with weights (see Figure \ref{fig:3star})
\begin{equation}\label{bubble3star}
\nu_1=\frac{2+a_1\epsilon}{\lambda},\quad\nu_2=\frac{1+a_2\epsilon}{\lambda},\quad\nu_3=\frac{1+a_3\epsilon}{\lambda},\quad a_1,a_2,a_3\in\RR.
\end{equation}
If $a_2=-1/2$ or $a_3=-1/2$, then $\nu_2=1$ or $\nu_3=1$ and the three-star is constructible, see Section \ref{sectcon}. It can be calculated to high order in $\epsilon$
by subtraction of the subdivergence from the bubble of weight $\nu_1$, see Section \ref{sectreg}. For general weights, appending an edge cannot be used to calculate
$\smallclaw(\nu_1,\nu_2,\nu_3)$.

Completion gives an extra edge from the internal vertex to $\infty$ of weight
$$
\nu_\infty=-\frac{(1+a_1+a_2+a_3)\epsilon}{\lambda}.
$$
The completed three-star also has edges between external vertices with weights $\nu_{z\infty}=-(1+a_3\epsilon)/\lambda$, $\nu_{01}=(-2-(1/2+a_1+a_2+a_3)\epsilon)\lambda$,
$\nu_{0\infty}=(1/2+a_2+a_3)\epsilon\lambda$, and $\nu_{1\infty}=(1+(1/2+a_1+a_3)\epsilon)\lambda$. These external edges are mere factors which pose no problem for the calculation
of $\smallclaw(\nu_1,\nu_2,\nu_3)$. We doubly transpose $0\leftrightarrow1$, $z\leftrightarrow\infty$ in the completed three-star.
The new vertex $z$ now has the weight $\nu_\infty$ which, using (\ref{eqconvolute}), we write as a chain of an edge of weight
$$
\nu_4=\frac{1-(1+a_1+a_2+a_3)\epsilon}\lambda
$$
(attached to the central vertex) and an edge of weight 1 (attached to $z$). We acquire a factor
$$
\frac{1}{c^{(\lambda)}_{\nu_4,1}}=-\Gamma(\lambda)\Big(1+\Big(\frac12+a_1+a_2+a_3\Big)\epsilon\Big)(1+a_1+a_2+a_3)\epsilon.
$$
Because the factor is of low degree $\epsilon$ we need to calculate the three-star with the appended edge of weight 1 to one order less than the original three-star.
The algorithm of appending an edge, see Section \ref{sectappedge}, does not increase the required order in $\epsilon$ and the terms from regularization of the logarithmic
singularity are trivial. We obtain the reduction
\begin{equation}\label{3starred}
\smallclaw(\nu_1,\nu_2,\nu_3)+\sO(\epsilon^{n+1})\longleftarrow\smallclaw(\nu_2,\nu_1,\nu_4)+\sO(\epsilon^n).
\end{equation}
A permutation of $0$ and $1$ on the right hand side leads to a bootstrap algorithm. From $\smallclaw(\nu_1,\nu_2,\nu_3)=0+\sO(\epsilon^{-1})$ we can construct
$\smallclaw(\nu_1,\nu_2,\nu_3)$ to any order in $\epsilon$ (subject to time and memory limitations).

Completion can also be useful for the calculation of two-point functions: We delete the two external legs and obtain a Feynman graph $G$ with two external vertices $z_0$ and $z_1$,
see Section \ref{sectper}. We set $z_0=0$ and $z_1=1$ without loss of information. The internal vertices have degree $(2n+4)/(n+1)$ whereas the external vertices $0$ and $1$ have
degree $(2n+4)/(n+1)-1=(n+3)/(n+1)$. Counting half-edges in $G$ gives
$$
\frac{2n+4}{n+1}V_G-2=2E_G,
$$
where $E_G$ and $V_G$ are the number of edges and vertices (internal or external) in $G$. From graph homology we get $E_G+1=h_G+V_G$, where $h_G$ is the number of independent
cycles in $G$ (the loop order). Combining both equations we get
\begin{equation}\label{VGhG}
V_G=(n+1)h_G.
\end{equation}
We complete the amputated two-point function by adding a vertex $\infty$ and an edge $01$
such that the completed graph $\overline G$ becomes $D/\lambda$-regular (i.e.\ every vertex has conformal degree $D/\lambda$). According to (\ref{wtinf}) the edges from the internal vertices
to $\infty$ have weight $\epsilon/(n+1)\lambda$. For the edges between $01$, $0\infty$, and $1\infty$ we get the weights
\begin{equation}\label{nus}
\nu_{01}=\frac{-1+\epsilon h_G/2}{\lambda},\qquad \nu_{0\infty}=\nu_{1\infty}=\frac{n+2-\epsilon(\frac{n-1}{n+1}+h_G)/2}{\lambda}.
\end{equation}
Using (\ref{VGhG}) it is easily checked that the completed graph $\overline G$ is $D/\lambda$-regular.
In completed Feynman periods we may forget labels: all choices of three vertices for $0$, $1$, $\infty$ give the same result \cite{numfunct}.
This is proved in general integer dimensions in \cite{gfe}. In non-integer dimensions it is a very well tested conjecture.

Note that the former vertex $\infty$ (now without label) has only two edges whose weights do not vanish in the limit $\epsilon\to0$. We will see in the next subsection that this property
can be utilized to derive an alternative representation for the period $P_G$.

\subsection{Integration by parts}\label{sectIBP}
Like in momentum space, integration by parts (IBP) uses Stokes' theorem to show that an integrand of the form $\sum_{\mu=1}^D\partial_\mu f_\mu$ is zero. In $2n+4-\epsilon$ dimensions
one obtains the conjecture that \cite{gfe}
\begin{gather}
0=~
(2\!-\!\nu_1\!-\!\sum\limits_{i=1}^N\nu_i) ~~
\begin{tikzpicture}[scale=1,baseline={([yshift=-0.7ex]current bounding box.center)}]
    \coordinate (vm) at (0, 0);
    \coordinate (v1) at (0, .7);
    \coordinate (v2) at (-.7, 0);
    \coordinate (v3) at (0,-.7);
    \draw[black!50] (vm) -- ([shift=(30:.3)]vm);
    \draw[black!50] (vm) -- ([shift=(0:.3)]vm);
    \draw[black!50] (vm) -- ([shift=(-30:.3)]vm);
    \draw (vm) -- node[right] {\tiny{$\nu_1\!+\!\frac{1}{\lambda}$}} (v1);
    \draw (vm) -- node[below] {\tiny{$\nu_2$}} (v2);
    \draw (vm) -- node[right] {\tiny{$\nu_3$}} (v3);
    \filldraw[black!20] (v1) -- ([shift=(60:.2)]v1) arc (60:120:.2) -- (v1);
    \draw (v1) -- ([shift=(60:.2)]v1);
    \draw (v1) -- ([shift=(120:.2)]v1);
    \filldraw[black!20] (v2) -- ([shift=(150:.2)]v2) arc (150:210:.2) -- (v2);
    \draw (v2) -- ([shift=(150:.2)]v2);
    \draw (v2) -- ([shift=(210:.2)]v2);
    \filldraw[black!20] (v3) -- ([shift=(240:.2)]v3) arc (240:300:.2) -- (v3);
    \draw (v3) -- ([shift=(240:.2)]v3);
    \draw (v3) -- ([shift=(300:.2)]v3);
    \filldraw (v1) circle (1.3pt);
    \filldraw (v2) circle (1.3pt);
    \filldraw (v3) circle (1.3pt);
    \filldraw (vm) circle (1.3pt);
\end{tikzpicture}
\hspace{-3ex}
+
\nu_2~~
\begin{tikzpicture}[scale=1,baseline={([yshift=-0.7ex]current bounding box.center)}]
    \coordinate (vm) at (0, 0);
    \coordinate (v1) at (0, .7);
    \coordinate (v2) at (-.7, 0);
    \coordinate (v3) at (0,-.7);
    \draw[black!50] (vm) -- ([shift=(30:.3)]vm);
    \draw[black!50] (vm) -- ([shift=(0:.3)]vm);
    \draw[black!50] (vm) -- ([shift=(-30:.3)]vm);
    \draw (vm) -- node[right] {\tiny{$\nu_1\!+\!\frac{1}{\lambda}$}} (v1);
    \draw (vm) -- node[below] {\tiny{$\nu_2\!+\!\frac{1}{\lambda}$}} (v2);
    \draw (vm) -- node[right] {\tiny{$\nu_3$}} (v3);
    \draw (v1) -- node[inner sep=1pt,above left] {\tiny{$-\frac{1}{\lambda}$}} (v2);
    \filldraw[black!20] (v1) -- ([shift=(60:.2)]v1) arc (60:120:.2) -- (v1);
    \draw (v1) -- ([shift=(60:.2)]v1);
    \draw (v1) -- ([shift=(120:.2)]v1);
    \filldraw[black!20] (v2) -- ([shift=(150:.2)]v2) arc (150:210:.2) -- (v2);
    \draw (v2) -- ([shift=(150:.2)]v2);
    \draw (v2) -- ([shift=(210:.2)]v2);
    \filldraw[black!20] (v3) -- ([shift=(240:.2)]v3) arc (240:300:.2) -- (v3);
    \draw (v3) -- ([shift=(240:.2)]v3);
    \draw (v3) -- ([shift=(300:.2)]v3);
    \filldraw (v1) circle (1.3pt);
    \filldraw (v2) circle (1.3pt);
    \filldraw (v3) circle (1.3pt);
    \filldraw (vm) circle (1.3pt);
\end{tikzpicture}
\hspace{-3ex}
-
\nu_2~~
\begin{tikzpicture}[scale=1,baseline={([yshift=-0.7ex]current bounding box.center)}]
    \coordinate (vm) at (0, 0);
    \coordinate (v1) at (0, .7);
    \coordinate (v2) at (-.7, 0);
    \coordinate (v3) at (0,-.7);
    \draw[black!50] (vm) -- ([shift=(30:.3)]vm);
    \draw[black!50] (vm) -- ([shift=(0:.3)]vm);
    \draw[black!50] (vm) -- ([shift=(-30:.3)]vm);
    \draw (vm) -- node[right] {\tiny{$\nu_1$}} (v1);
    \draw (vm) -- node[below] {\tiny{$\nu_2\!+\!\frac{1}{\lambda}$}} (v2);
    \draw (vm) -- node[right] {\tiny{$\nu_3$}} (v3);
    \filldraw[black!20] (v1) -- ([shift=(60:.2)]v1) arc (60:120:.2) -- (v1);
    \draw (v1) -- ([shift=(60:.2)]v1);
    \draw (v1) -- ([shift=(120:.2)]v1);
    \filldraw[black!20] (v2) -- ([shift=(150:.2)]v2) arc (150:210:.2) -- (v2);
    \draw (v2) -- ([shift=(150:.2)]v2);
    \draw (v2) -- ([shift=(210:.2)]v2);
    \filldraw[black!20] (v3) -- ([shift=(240:.2)]v3) arc (240:300:.2) -- (v3);
    \draw (v3) -- ([shift=(240:.2)]v3);
    \draw (v3) -- ([shift=(300:.2)]v3);
    \filldraw (v1) circle (1.3pt);
    \filldraw (v2) circle (1.3pt);
    \filldraw (v3) circle (1.3pt);
    \filldraw (vm) circle (1.3pt);
\end{tikzpicture}
+
\nu_3
\hspace{-2ex}
\begin{tikzpicture}[scale=1,baseline={([yshift=-0.7ex]current bounding box.center)}]
    \coordinate (vm) at (0, 0);
    \coordinate (v1) at (0, .7);
    \coordinate (v2) at (-.7, 0);
    \coordinate (v3) at (0,-.7);
    \draw[black!50] (vm) -- ([shift=(30:.3)]vm);
    \draw[black!50] (vm) -- ([shift=(0:.3)]vm);
    \draw[black!50] (vm) -- ([shift=(-30:.3)]vm);
    \draw (vm) -- node[right] {\tiny{$\nu_1\!+\!\frac{1}{\lambda}$}} (v1);
    \draw (vm) -- node[below] {\tiny{$\nu_2$}} (v2);
    \draw (vm) -- node[right] {\tiny{$\nu_3\!+\!\frac{1}{\lambda}$}} (v3);
    \draw (v1) .. controls (-1.4,.8) and (-1.4,-.8) .. (v3) node[pos=.3,inner sep=1pt,above left] {\tiny{$-\frac{1}{\lambda}$}} (v2);
    \filldraw[black!20] (v1) -- ([shift=(60:.2)]v1) arc (60:120:.2) -- (v1);
    \draw (v1) -- ([shift=(60:.2)]v1);
    \draw (v1) -- ([shift=(120:.2)]v1);
    \filldraw[black!20] (v2) -- ([shift=(150:.2)]v2) arc (150:210:.2) -- (v2);
    \draw (v2) -- ([shift=(150:.2)]v2);
    \draw (v2) -- ([shift=(210:.2)]v2);
    \filldraw[black!20] (v3) -- ([shift=(240:.2)]v3) arc (240:300:.2) -- (v3);
    \draw (v3) -- ([shift=(240:.2)]v3);
    \draw (v3) -- ([shift=(300:.2)]v3);
    \filldraw (v1) circle (1.3pt);
    \filldraw (v2) circle (1.3pt);
    \filldraw (v3) circle (1.3pt);
    \filldraw (vm) circle (1.3pt);
\end{tikzpicture}
\hspace{-3ex}
-
\nu_3~~
\begin{tikzpicture}[scale=1,baseline={([yshift=-0.7ex]current bounding box.center)}]
    \coordinate (vm) at (0, 0);
    \coordinate (v1) at (0, .7);
    \coordinate (v2) at (-.7, 0);
    \coordinate (v3) at (0,-.7);
    \draw[black!50] (vm) -- ([shift=(30:.3)]vm);
    \draw[black!50] (vm) -- ([shift=(0:.3)]vm);
    \draw[black!50] (vm) -- ([shift=(-30:.3)]vm);
    \draw (vm) -- node[right] {\tiny{$\nu_1$}} (v1);
    \draw (vm) -- node[below] {\tiny{$\nu_2$}} (v2);
    \draw (vm) -- node[right] {\tiny{$\nu_3\!+\!\frac{1}{\lambda}$}} (v3);
    \filldraw[black!20] (v1) -- ([shift=(60:.2)]v1) arc (60:120:.2) -- (v1);
    \draw (v1) -- ([shift=(60:.2)]v1);
    \draw (v1) -- ([shift=(120:.2)]v1);
    \filldraw[black!20] (v2) -- ([shift=(150:.2)]v2) arc (150:210:.2) -- (v2);
    \draw (v2) -- ([shift=(150:.2)]v2);
    \draw (v2) -- ([shift=(210:.2)]v2);
    \filldraw[black!20] (v3) -- ([shift=(240:.2)]v3) arc (240:300:.2) -- (v3);
    \draw (v3) -- ([shift=(240:.2)]v3);
    \draw (v3) -- ([shift=(300:.2)]v3);
    \filldraw (v1) circle (1.3pt);
    \filldraw (v2) circle (1.3pt);
    \filldraw (v3) circle (1.3pt);
    \filldraw (vm) circle (1.3pt);
\end{tikzpicture}
\hspace{-3ex}
+~\ldots,
\label{eqibpgeneral}
\end{gather}
where the shaded cones at the external vertices indicate that the graphs can be substituted into an ambient graphical function or Feynman period.
The $\nu_i$, $i=1,\ldots,N$, are the indicated edge weights.

After completion, Identity (\ref{eqibpgeneral}) has a particularly symmetric form. The central vertex acquires an extra edge that connects to $\infty$.
Accordingly, we obtain an extra weight $\nu_{N+1}$. The full set of weights is restricted by the identity
$$
\sum_{i=1}^{N+1}\nu_i=2.
$$
From a star with $N+1$ edges of weights $\nu_1,\ldots,\nu_{N+1}$ that connect the central vertex $x$ to the external vertices $z_1,\ldots,z_{N+1}$
we construct graphs $g_{ij}=g_{ji}$ for $i\neq j$ by adding triangles $xz_i$, $xz_j$, $z_iz_j$ with weights $\nu_{xz_i}=\nu_{xz_j}=1/\lambda$ and $\nu_{z_iz_j}=-1/\lambda$.
Note that corresponding vertices in all graphs $g_{ij}$ have equal total weight. In particular, the weight of the central vertex $x$ is $D/\lambda$.
We also consider the graphs $g_{ij}$ as insertions into a larger graph.
To do this we fix a completed graphical function or a completed Feynman period $\Gg$. We choose an internal vertex $x$ in $\Gg$ and label the $N+1$ adjacent
(internal or external) vertices by $z_1,\ldots,z_{N+1}$. For $a\neq b$ in $1,\ldots,N+1$ we define $\Gg^{ab}=\Gg^{ba}$ as the graph $\Gg$ with an additional triangle
$xz_a$, $xz_b$, $z_az_b$ of weights $\nu_{xz_a}=\nu_{xz_b}=-1/\lambda$ and $\nu_{z_az_b}=1/\lambda$. Note that the weights in the triangle are opposite to those
in $g_{ij}$ (hence the superscript). For a fixed graph $\Gg^{ab}$ we define a set of completed graphs $\Gg^{ab}_{ij}$ for $i\neq j$ with triangle insertions
at $x$, $z_i$, $z_j$ of weights $\nu_{xz_i}=\nu_{xz_j}=1/\lambda$ and $\nu_{z_iz_j}=-1/\lambda$. In particular, we have $\Gg^{ab}_{ab}=\Gg$.

In this completed setup for graphical functions or Feynman periods, Identity (\ref{eqibpgeneral}) reduces to \cite{gfe}
\begin{equation}\label{IBPcompleted}
\sum_{\genfrac{}{}{0pt}{}{j=1}{j\neq i}}^{N+1}\nu_jf_{\Gg^{ab}_{ij}}(z)=:F_{\Gg^{ab}}(z)\quad\text{or}\quad\sum_{\genfrac{}{}{0pt}{}{j=1}{j\neq i}}^{N+1}\nu_jP_{\Gg^{ab}_{ij}}=:F_{\Gg^{ab}},
\end{equation}
where $F_{\Gg^{ab}}$ does not depend on $i$. The graphical function (or the Feynman period) of a completed graph is defined as the graphical function (or Feynman period)
after the removal of $\infty$.

\begin{figure}
\centering
\includegraphics{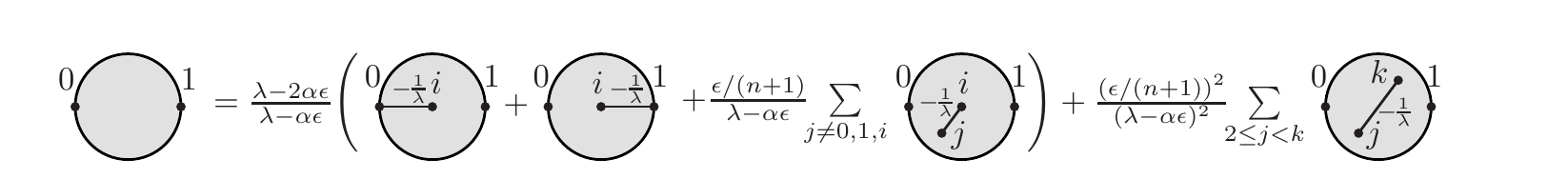}
\caption{A formula for the Feynman period of an amputated two-point function ($\alpha=\frac{h_G}2-\frac1{n+1}$).}
\label{fig:IBP2pt}
\end{figure}

We exemplify the completed IBP relation (\ref{IBPcompleted}) in the setup of the previous subsection.
We completed the amputated two-point function $G$ with external vertices $0$ and $1$ by adding $\infty$ yielding the completed Feynman period $\Gg$.
For notation we keep the labels $0,1,\infty$ in $\Gg$ although $\Gg$ needs no distinction between internal and external vertices.
We use $x=\infty$ as central IBP vertex and consider $\Gg^{01}$. From (\ref{nus}) we obtain
\begin{equation}\label{nu0}
\nu_0=\nu_1=\nu_{0\infty}-\frac1\lambda=1-\frac{\alpha\epsilon}{\lambda},\qquad\text{with}\qquad\alpha=\frac{h_G}{2}-\frac1{n+1}.
\end{equation}
The completed IBP relation (\ref{IBPcompleted}) for $i=0,1$ gives
$$
\Big(1-\frac{\alpha\epsilon}\lambda\Big)P_\Gg+\frac{\epsilon}{(n+1)\lambda}\sum_{j=2}^{V_G-1}P_{\Gg^{01}_{ij}}=F_{\Gg^{01}},
$$
where we labeled the $V_G-2$ internal vertices of the two-point graph $G$ by $2,\ldots,V_G-1$. Subtraction of the two identities for $i=0$ and $i=1$ gives
$$
\sum_{j=2}^{V_G-1}P_{\Gg^{01}_{0j}}=\sum_{j=2}^{V_G-1}P_{\Gg^{01}_{1j}}.
$$
For $i\geq2$ we obtain
$$
\Big(1-\frac{\alpha\epsilon}\lambda\Big)(P_{\Gg^{01}_{0i}}+P_{\Gg^{01}_{1i}})+\frac{\epsilon}{(n+1)\lambda}\sum_{j\neq0,1,i}P_{\Gg^{01}_{ij}}=F_{\Gg^{01}}.
$$
We sum over all internal vertices $i=2,\ldots,V_G-1$, use the previous equation and substitute $\sum_jP_{\Gg^{01}_{1j}}$ into the first equation for $i=1$ to obtain
$$
\Big(1-\frac{\alpha\epsilon}\lambda\Big)P_\Gg=\frac{\epsilon^2}{(n+1)^2\lambda(\lambda-\alpha\epsilon)}\sum_{2\leq j<k\leq V_G-1}P_{\Gg^{01}_{jk}}+
\frac{\lambda-2\alpha\epsilon}{\lambda-\alpha\epsilon}F_{\Gg^{01}}.
$$
We take $F_{\Gg^{01}}$ from the previous equation and get for any $i\in\{2,\ldots,V_G-1\}$ the following alternative representation of the two-point period $P_\Gg$,
\begin{equation}\label{twopointeq}
P_\Gg=\frac{\lambda-2\alpha\epsilon}{\lambda-\alpha\epsilon}\Big(P_{\Gg^{01}_{0i}}+P_{\Gg^{01}_{1i}}+\frac{\epsilon}{(n+1)(\lambda-\alpha\epsilon)}\sum_{j\neq0,1,i}P_{\Gg^{01}_{ij}}\Big)
+\frac{\epsilon^2}{(n+1)^2(\lambda-\alpha\epsilon)^2}\sum_{2\leq j<k\leq V_G-1}P_{\Gg^{01}_{jk}}.
\end{equation}
The  Feynman periods on the right hand side of (\ref{twopointeq}) are completed. We may label the central vertex in the IBP relation by $\infty$
(although it was internal for deriving the IBP relation) and stick to the original labels 0 and 1 to obtain the de-completed relation that is depicted in Figure \ref{fig:IBP2pt}.

Note that a priori the right hand side of (\ref{twopointeq}) is not simpler than the left hand side. The Laurent series' of $P_{\Gg^{01}_{ij}}$ and $P_{\Gg^{01}_{jk}}$
are only needed to lower order in $\epsilon$. These terms are typically easier to calculate. However, the calculation of the Feynman periods $P_{\overline{G}_{0j}}$ and
$P_{\overline{G}_{1j}}$ may be as hard as the original period $P_{\overline G}$. The benefit of (\ref{twopointeq}) is the freedom to choose the label $i$.
If $P_{\overline G}$ is inaccessible, it happens quite frequently that there exist an $i$ so that the right hand side of (\ref{twopointeq}) can be computed
(although the right hand side cannot be computed of most other choices of $i$).

\begin{figure}
\centering
\includegraphics{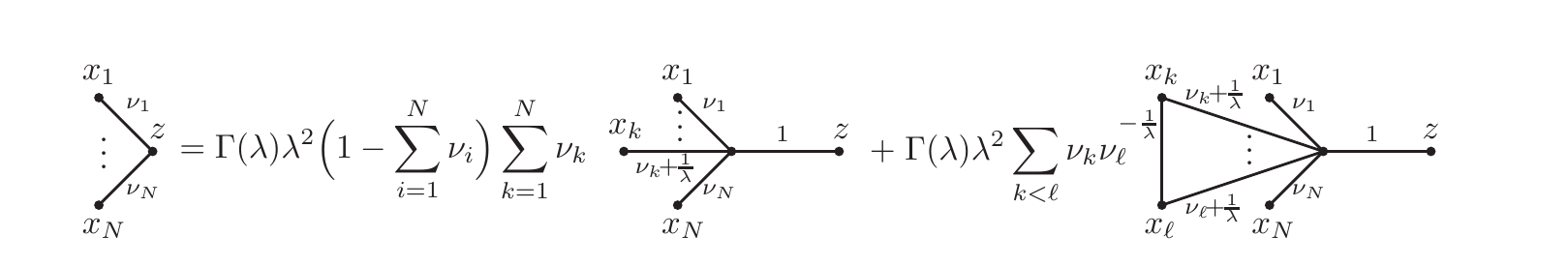}
\caption{An IBP relation with a contraction on the left hand side ($\nu_1,\ldots,\nu_N\neq1$).}
\label{fig:IBP2}
\end{figure}

\begin{figure}
\centering
\includegraphics{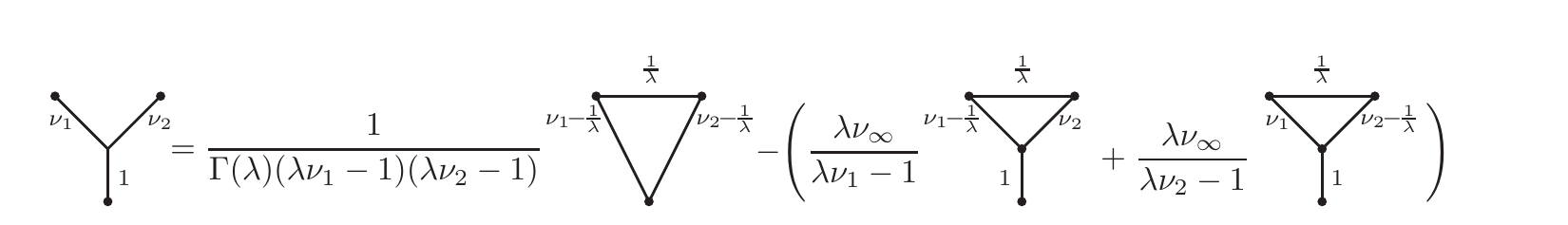}
\caption{The Y$\nabla\overline{\mathrm{Y}}\,\overline{\mathrm{Y}}$-identity, where $\nu_\infty=1+\frac2\lambda-\nu_1-\nu_2$.
This identity is particularly useful for $\nu_1=\nu_2=1$ in $6-\epsilon$ dimensions.}
\label{fig:YDYY}
\end{figure}

A special case arises if the differential operator produces a Dirac delta-function by
\begin{align}
\label{eq:propagator}
\Box_x\frac{1}{||x-y|||^{2\lambda}} =-\frac{4}{\Gamma(\lambda)}\delta(x-y).
\end{align}
Then, a term is generated where an edge is contracted. The general case in Figure \ref{fig:IBP2} can be useful if all edges $\nu_1,\ldots,\nu_N$ have weight zero
in the limit $\epsilon=0$. In this case, the terms on the right hand side need lower orders in the $\epsilon$-expansion. The case $N=1$ was used in Section \ref{sectcom} for the calculation
of the general three-star $\smallclaw(\nu_1,\nu_2,\nu_3)$ with a bubble in four dimensions, see (\ref{bubble3star}). Lower orders in the $\epsilon$-expansion also suffice
if one edge has non-zero weight for $\epsilon=0$ but the sum of all weights is $1+\sO(\epsilon)$.

An important case is also $N=2$, see Figure \ref{fig:YDYY}. The contraction gives rise to the triangle on the right hand side. This Y$\nabla\overline{\mathrm{Y}}\,\overline{\mathrm{Y}}$-identity is a main tool in $6-\epsilon$ dimensions for weights $\nu_1=\nu_2=1$ because the coefficient
$\epsilon/(2-\epsilon)$ of the two rightmost graphs suppresses one order in $\epsilon$. Note that in six dimensions the Y$\nabla\overline{\mathrm{Y}}\,\overline{\mathrm{Y}}$-identity
reduces for $\epsilon=0$ to a star-triangle identity which eliminates a vast number of internal vertices (and hence integrations) in $\phi^3$ theory \cite{phi3,DY}.

The Y$\nabla\overline{\mathrm{Y}}\,\overline{\mathrm{Y}}$-identity may also be used after completion (and de-completion), see Section \ref{sectcom}.
From the $\phi^3$ three-point function we obtain an effective two-point function by taking one external vertex to infinity (thus eliminating the leg), see Section \ref{sectren}.
After the amputation of the two remaining external legs we obtain a two-point function with external vertices $0$ and $1$.
Completion connects the external vertices $0$ and $1$ to $\infty$ by edges of weight $1-(h_G-1)\epsilon/2\lambda$, see (\ref{nu0}). If one deletes the vertex $0$ (or $1$) as new vertex $\infty$,
then one can use the Y$\nabla\overline{\mathrm{Y}}\,\overline{\mathrm{Y}}$-identity at vertex $1$ (or $0$).

\section{Approximation}\label{sectapp}
Many relations between graphical functions are only valid up to a certain power $\epsilon^N$. Such relations can be used to approximate a graphical function by a sum of
graphical functions. Beyond the order $\epsilon^N$ the approximation fails. These approximations typically work locally on the level of the integrand.
One replaces a certain subgraph in the graph $G$ by a sum of simpler subgraphs with the same evaluation up to order $\epsilon^N$.
In the graphical function the subgraph corresponds to a set of internal vertices and the attached edges. If the graphical function of $G$ is singular in integer dimensions,
$\epsilon$ is a regulator and the singularity structure must not be changed by the approximation. The singularity structure can be identified by power-counting of subgraphs
(see e.g.\ \cite{gfe}). If a certain sub-structure $g$ of $G$ has a scaling weight $N_g$ which characterizes a logarithmic divergence at $\epsilon=0$, then every term in the
approximation needs to have the exact same scaling weight for $\epsilon\neq0$. Examples of this type are in Section \ref{sectqu}, see Figures \ref{fig:app1} and \ref{fig:app2}.

\subsection{Expansion}\label{sectexp}
Assume the graph $G$ of a graphical function $f_G(z)$ has a set of edges $\sE$ such that $f_G(z)$ simplifies if one changes the coefficient of $\epsilon$ in any of the edges in $\sE$.
Concretely, assume that edges $e\in\sE$ have weight $\nu_e=(n_e+a_e\epsilon)/\lambda$, see (\ref{nua}), and $f_G(z)$ becomes accessible if the weights of some edges in $\sE$
change to $\nu_e'=(n_e+a_e'\epsilon)/\lambda$. Assume we want to know the lowest $N+1$ Laurent coefficients of $\epsilon^L,\ldots,\epsilon^{L+N}$ in $f_G(z)$,
where $L$ is the low-degree of $f_G(z)$ in $\epsilon$. If $N<|\sE|$ and no singular subgraph in integer dimensions (for $\epsilon=0$) uses any of the edges in $\sE$,
then we can approximate the integrand by a linear combination of $N+1$ graphical functions with modified weights $a_e'$.

A standard example is the case $n_e=a_e'=0$. By completion, from graphs with $(2n+4)/(n+1)$ edges of weight 1 at each internal vertex ($n=0,1$) one obtains graphical functions
with $V^{\mathrm{int}}$ edges of weight $\epsilon/(n+1)\lambda$, see Section \ref{sectcom}. In this case $a_e=1/(n+1)$ and the graph simplifies (loses edges) by the transition to $a_e'=0$.
We obtain the following identity for $N<|\sE|$,
\begin{equation}\label{eqexpand}
\prod_{e\in\sE}||x_e||^{-2a_e\epsilon}=\sum_{k=0}^N(-1)^{N-k}\genfrac(){0pt}{}{|\sE|-k-1}{N-k}\sum_{\sE_k\subset\sE}\prod_{e\in\sE_k}||x_e||^{-2a_e\epsilon}+\sO(\epsilon^{N+1}),
\end{equation}
where $x_e=x_i-x_j$ if the edge $e$ is $x_ix_j$, and $\sE_k$ is a subset of $\sE$ with $k$ elements. If this identity is used in the integrand of the graphical function $f_G(z)$
after a double transposition of external vertices, one obtains an approximation for the lowest $N+1$ coefficients using $\leq N$ edges of weight $a_e\epsilon/\lambda$.

Another application of approximation is the conversion of an edge of weight $a\epsilon/\lambda$ into a sum of two edges of weight $-\epsilon/2\lambda$ and $0$.
This approximation only holds up to order $\epsilon$,
$$
||x_e||^{-2a\epsilon}=-2a||x_e||^\epsilon+2a+1+\sO(\epsilon^2).
$$
The benefit of this approximation is that in four dimensions an edge of weight $-\epsilon/2\lambda$ can be converted into a chain of two edges of weight $1$, see (\ref{eqconvolute}).
With $c^{(\lambda)}_{1,1}=-2/\epsilon\Gamma(\lambda)$ we obtain,
\begin{equation}\label{epsilonred}
\begin{tikzpicture}[baseline={([yshift=-2.5ex]current bounding box.center)}]
    \coordinate (v1) at (0,0);
    \coordinate (v3) at (1.2,0);
    \filldraw[black!20] (v1) -- ([shift=(150:.2)]v1) arc (150:210:.2) -- (v1);
    \draw (v1) -- ([shift=(150:.2)]v1);
    \draw (v1) -- ([shift=(210:.2)]v1);
    \filldraw[black!20] (v3) -- ([shift=(30:.2)]v3) arc (30:-30:.2) -- (v3);
    \draw (v3) -- ([shift=(30:.2)]v3);
    \draw (v3) -- ([shift=(-30:.2)]v3);
    \filldraw (v1) circle (1.3pt);
    \filldraw (v3) circle (1.3pt);
    \draw (v1) -- node[above] {$\frac{a\epsilon}{\lambda}$} (v3);
\end{tikzpicture}
\quad
=\quad \Gamma(\lambda)a\epsilon\;
\begin{tikzpicture}[baseline={([yshift=-2.5ex]current bounding box.center)}]
    \coordinate (v1) at (0,0);
    \coordinate (v2) at (0.6,0);
    \coordinate (v3) at (1.2,0);
    \filldraw[black!20] (v1) -- ([shift=(150:.2)]v1) arc (150:210:.2) -- (v1);
    \draw (v1) -- ([shift=(150:.2)]v1);
    \draw (v1) -- ([shift=(210:.2)]v1);
    \filldraw[black!20] (v3) -- ([shift=(30:.2)]v3) arc (30:-30:.2) -- (v3);
    \draw (v3) -- ([shift=(30:.2)]v3);
    \draw (v3) -- ([shift=(-30:.2)]v3);
    \filldraw (v1) circle (1.3pt);
    \filldraw (v2) circle (1.3pt);
    \filldraw (v3) circle (1.3pt);

    \draw[white] (v1) -- node[above] {$\phantom{\nu_1+\nu_2-\frac{\lambda+1}{\lambda}}$} (v3);
    \draw (v1) -- node[above] {$1$} (v2);
    \draw (v2) -- node[above] {$1$} (v3);
\end{tikzpicture}
\;
+\;\;(2a+1)
\begin{tikzpicture}[baseline={([yshift=-2.5ex]current bounding box.center)}]
    \coordinate (v1) at (0,0);
    \coordinate (v3) at (1.2,0);
    \filldraw[black!20] (v1) -- ([shift=(150:.2)]v1) arc (150:210:.2) -- (v1);
    \draw (v1) -- ([shift=(150:.2)]v1);
    \draw (v1) -- ([shift=(210:.2)]v1);
    \filldraw[black!20] (v3) -- ([shift=(30:.2)]v3) arc (30:-30:.2) -- (v3);
    \draw (v3) -- ([shift=(30:.2)]v3);
    \draw (v3) -- ([shift=(-30:.2)]v3);
    \draw[white] (v1) -- node[above] {$\phantom{\nu_1+\nu_2-\frac{\lambda+1}{\lambda}}$} (v3);
    \filldraw (v1) circle (1.3pt);
    \filldraw (v3) circle (1.3pt);
\end{tikzpicture}\;+\;\sO(\epsilon^2).
\end{equation}
If the edge on the left hand side is the only edge that connects to $z$, the graphical function can be reduced by double use of the appending an edge algorithm, see Section \ref{sectappedge}.

\begin{figure}
\centering
\includegraphics{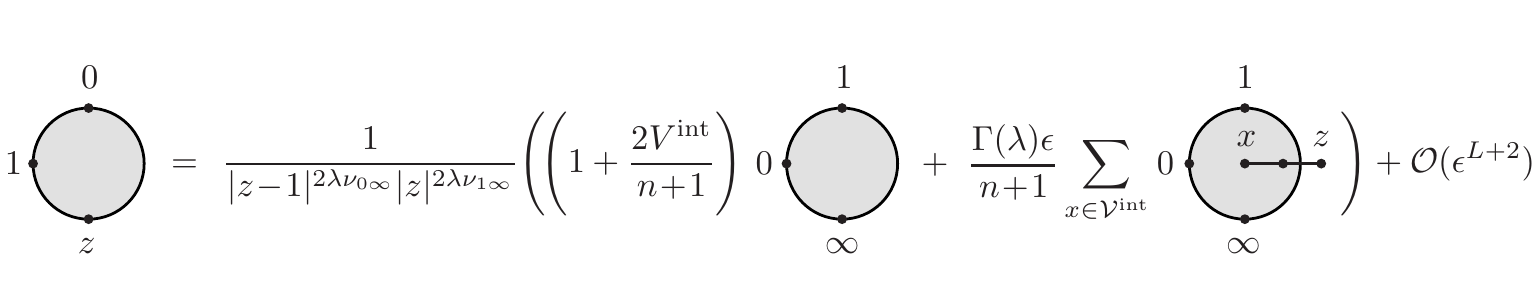}
\caption{A formula for the first two orders in $\epsilon$ of graphical functions with $\frac{2n+4}{n+1}$-valent internal vertices in $2n+4-\epsilon$ dimensions ($n=0,1$).
On the right hand side the vertex $\infty$ has to be deleted, see (\ref{confexp}).}
\label{fig:expand}
\end{figure}

\begin{figure}
\centering
\includegraphics{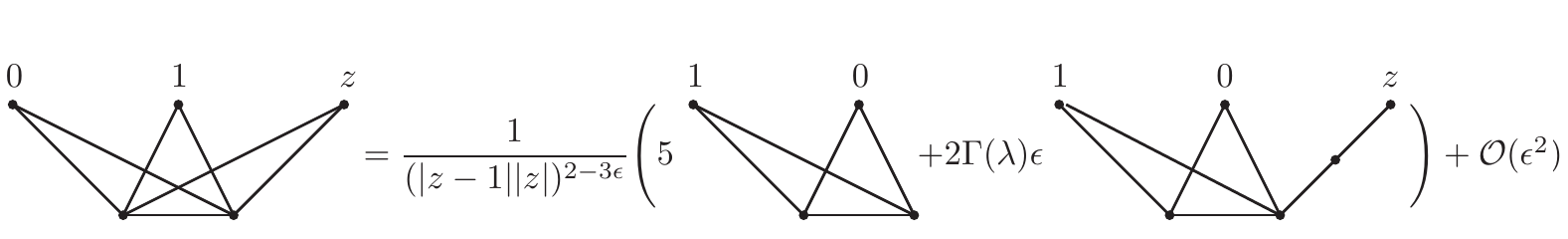}
\caption{A four-dimensional example for (\ref{confexp}), see Figure \ref{fig:expand}. The rightmost graph represents the sum of two isomorphic graphs.}
\label{fig:expandex}
\end{figure}

Finally, we combine completion with (\ref{eqexpand}) and (\ref{epsilonred}) to derive an identity for the Laurent expansion of a graphical function $f_G(z)$ with edges of weight 1
in $\phi^3$ or $\phi^4$ theory. Assume the $V^{\mathrm{int}}$ internal vertices of $G$ have $(2n+4)/(n+1)$ edges of weight 1 for $n=0,1$, so that internal vertices connect to $\infty$
with edges of weight $\epsilon/(n+1)\lambda$, see (\ref{wtinf}). Moreover, the vertex $\infty$ connects to $0$ and $1$ with
$$
\nu_{0\infty}=\frac{-N_0+N_1+N_z}2-\frac{V^{\mathrm{int}}\epsilon}{2(n+1)\lambda},\qquad \nu_{1\infty}=\frac{N_0-N_1+N_z}2-\frac{V^{\mathrm{int}}\epsilon}{2(n+1)\lambda},
$$
where $N_0$, $N_1$, $N_z$ are the total weights of the external vertices $0$, $1$, $z$, respectively. In the completed graph there also exist weighted edges $01$ and $z\infty$.
Their weights, however, are insignificant in the following. We use the double transposition $0\leftrightarrow1$, $z\leftrightarrow\infty$ to obtain after de-completion
$$
f_G(z)=\frac{1}{((z-1)(\zz-1))^{\lambda\nu_{0\infty}}(z\zz)^{\lambda\nu_{1\infty}}}f_{G'}(z),
$$
where the graph $G'$ is $G$ with $0\leftrightarrow1$, the vertex $z$ deleted, and a new vertex $z$ that connects to all internal vertices with edges of weight $\epsilon/(n+1)\lambda$.
Let $L$ be the low degree of the Laurent expansion of $f_G(z)$. We use (\ref{eqexpand}) for $N=1$ to expand $f_{G'}(z)$ to order $\epsilon^{L+1}$ (keeping the lowest two orders in $\epsilon$).
The expansion is $1-V^{\mathrm{int}}$ times $G'$ with no edges attached to $z$ plus a sum over graphs with single edges between any fixed internal vertex and $z$.
Using (\ref{epsilonred}) for $a=1/(n+1)$ we obtain
\begin{equation}\label{confexp}
f_G(z)=\frac{1}{((z-1)(\zz-1))^{\lambda\nu_{0\infty}}(z\zz)^{\lambda\nu_{1\infty}}}\Big(\Big(1+\frac{2V^{\mathrm{int}}}{n+1}\Big)f_{G_0}(z)+
\frac{\Gamma(\lambda)\epsilon}{n+1}\sum_{x\in\sV^{\mathrm{int}}}f_{G_x}(z)\Big)+\sO(\epsilon^{L+2}),
\end{equation}
where $G_0$ is the graph $G$ with $0\leftrightarrow1$ and deleted vertex $z$. The graph $G_x$ is $G_0$ with a chain of two vertices of weight $1$ between the internal vertex $x$ and a new
external vertex $z$. With (\ref{confexp}) the lowest two orders in $\epsilon$ of the left hand side can often be calculated. This identity is useful to calculate the top loop order contribution
to the beta-function. It is depicted in Figure \ref{fig:expand} with a special case in Figure \ref{fig:expandex}.

\subsection{Three-vertex cuts}\label{sectqu}
\begin{figure}
\centering
\includegraphics{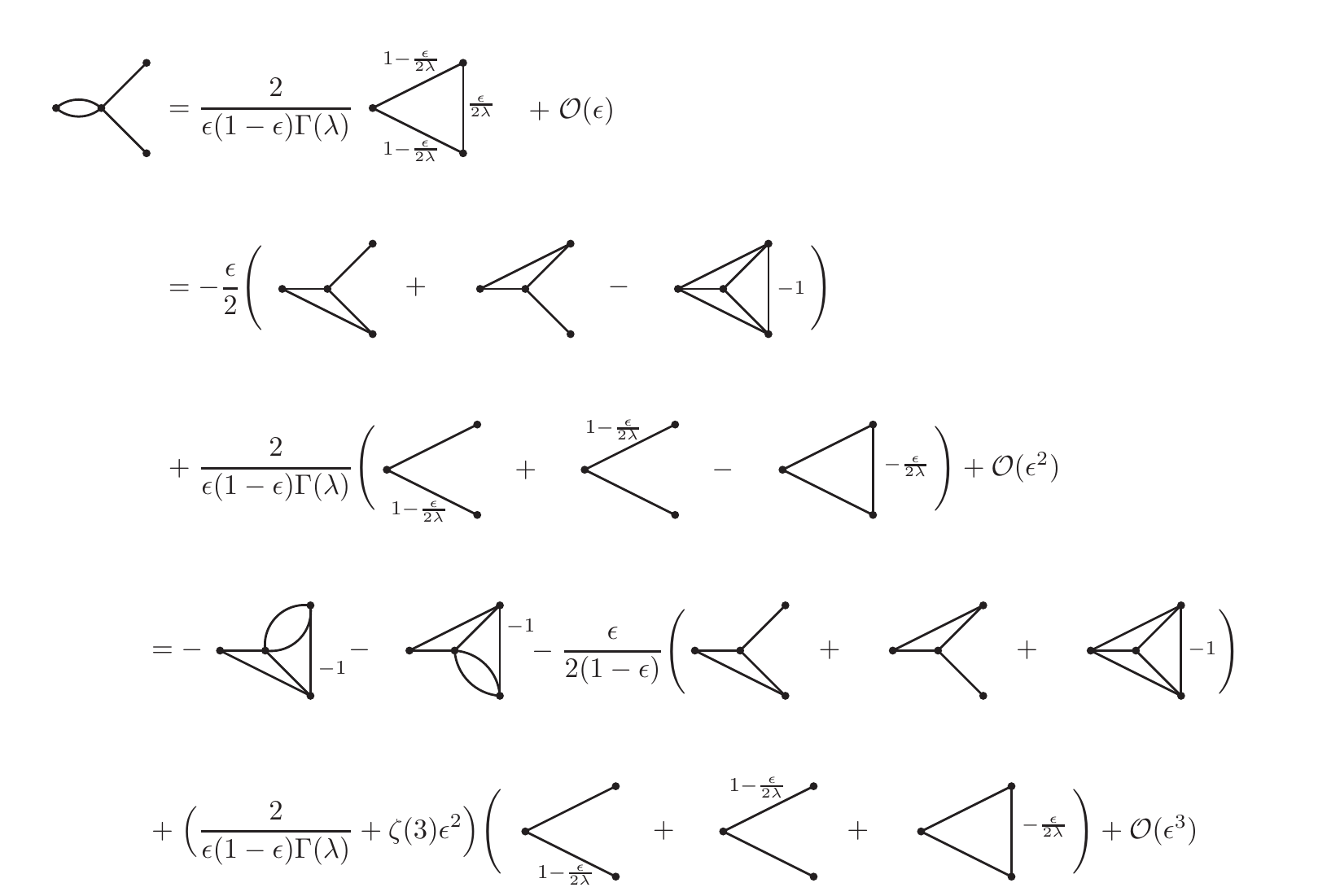}
\caption{Approximations of the three-star $\smallclaw(2,1,1)$ in four dimensions. Edges with no labels have weight 1.}
\label{fig:app1}
\end{figure}

\begin{figure}
\centering
\includegraphics{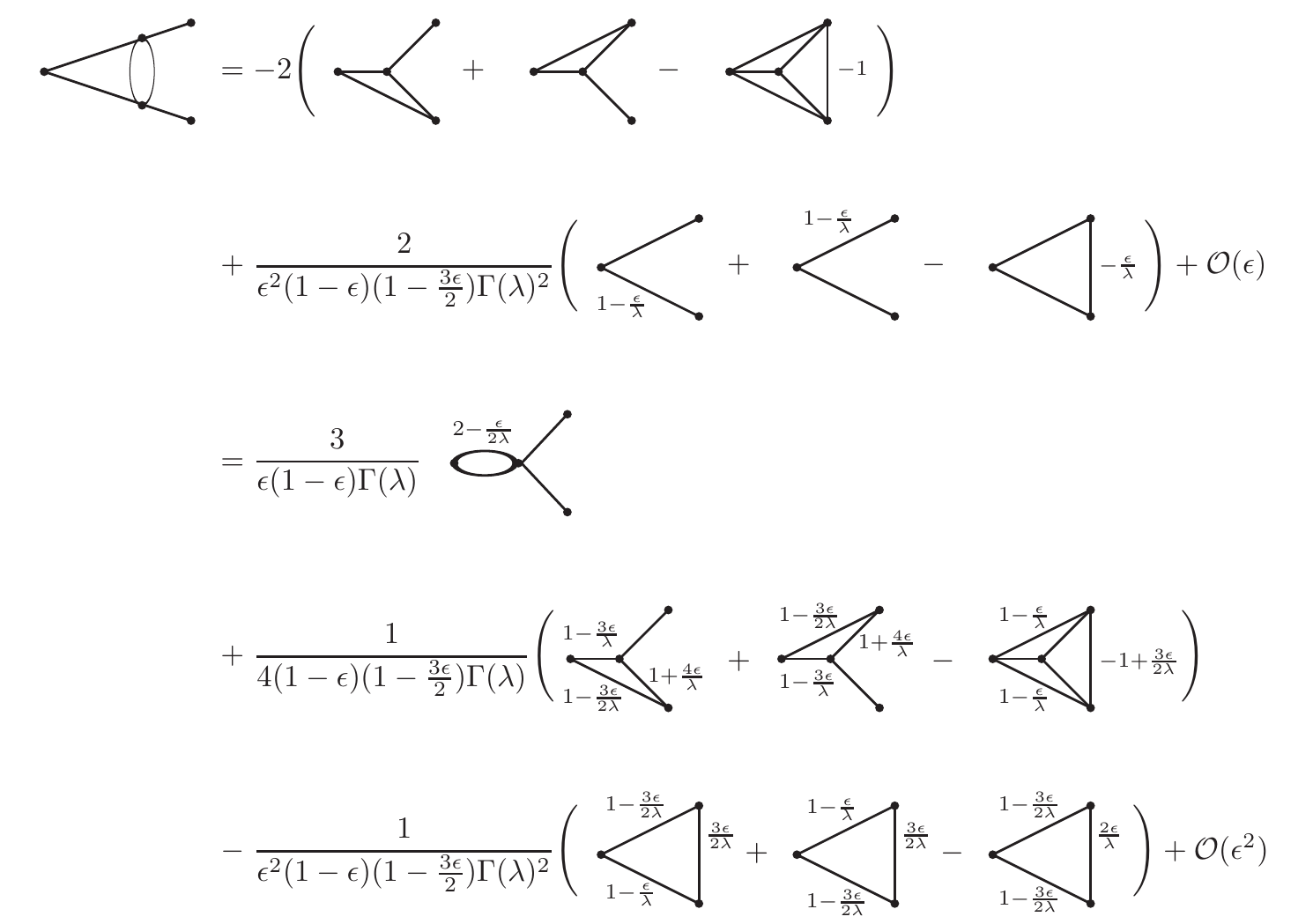}
\caption{Approximations of Dunce's cap in four dimensions. Edges with no labels have weight 1.}
\label{fig:app2}
\end{figure}

Many graphical functions have three-vertex cuts: they split after the removal of three vertices. This happens at each vertex in $\phi^3$ theory and at each vertex
with a bubble in $\phi^4$ theory. In practice, one often also finds less trivial three-vertex cuts in relevant graphical functions.

If a graph $G$ has a three-vertex cut along $x_1$, $x_2$, $x_3$, one can consider the vertices of the cut as external vertices of a smaller graphical function $g$. If $f_g(z)$ can be
approximated by a sum of simpler graphical functions up to some order in $\epsilon$, then the Feynman integral $A_g(x_1,x_2,x_3)$ is approximated by the sum of the simpler Feynman integrals.
From (\ref{AGinvs}) we obtain the condition that in each graph of the approximation the edge $01$ acquires a weight such that the scaling $N_g$ of $g$ is preserved, see (\ref{NGdef}).
Figures \ref{fig:app1} and \ref{fig:app2} are important examples for such approximations to various order in $\epsilon$ in four dimensions. We may replace $g$ in $G$ by the approximation
if the order in $\epsilon$ suffices and if no singular (by power-counting) subgraph of $G$ uses edges in $g$.

The latter condition can be relaxed if the power-counting of the singularity is preserved by each graph in the approximation. As an example of this (rather rare) situation,
consider the first approximation of the bubble three-star in Figure \ref{fig:app1}. Assume the left hand side is substituted into an ambient graph with an edge
of weight 1 from the left to the top vertex. In this case one obtains a logarithmic singularity in the shape of Dunce's cap (spanned by the left, top, and central vertex).
The cap contains the bubble, the central vertex, and the top edge of the graph $g$ on the left hand side in Figure \ref{fig:app1}. The weight of this chain is
$3-(\lambda+1)/\lambda=1-\epsilon/2\lambda$, see (\ref{eqconvolute}). The left-top edge on the sole graph in the first approximation on the right hand side reproduces this weight.
So, the first approximation in Figure \ref{fig:app1} can be used in the presence of a left-top edge of weight 1 in the ambient graph. The other approximations
in Figure \ref{fig:app1} cannot be used in this case because there exist graphs in the approximation with left-top edges of weight 1 ($\neq1-\epsilon/2\lambda$).

Some of the identities in Figures \ref{fig:app1} and \ref{fig:app2} generalize to weights $(1+a_e\epsilon)/\lambda$, see (\ref{nua}), which makes them significantly
more powerful. In {\tt HyperlogProcedures} \cite{Shlog} many more three-vertex approximations are implemented. A systematic and comprehensive search for three-vertex approximations
is still missing.

Approximations of three-vertex cuts are an important tool for the calculation of renormalization functions in four-dimensional $\phi^4$ theory. There exist analogous
three-vertex cut approximations in six-dimensional $\phi^3$ theory.
As a rule of thumb, however, IBP (see Section \ref{sectIBP}) is more important in $\phi^3$ theory whereas three-vertex cut approximations are more important in $\phi^4$ theory.

\section{Subtraction of subdivergences by rerouting}\label{sectrerou}

\begin{figure}
\centering
\includegraphics{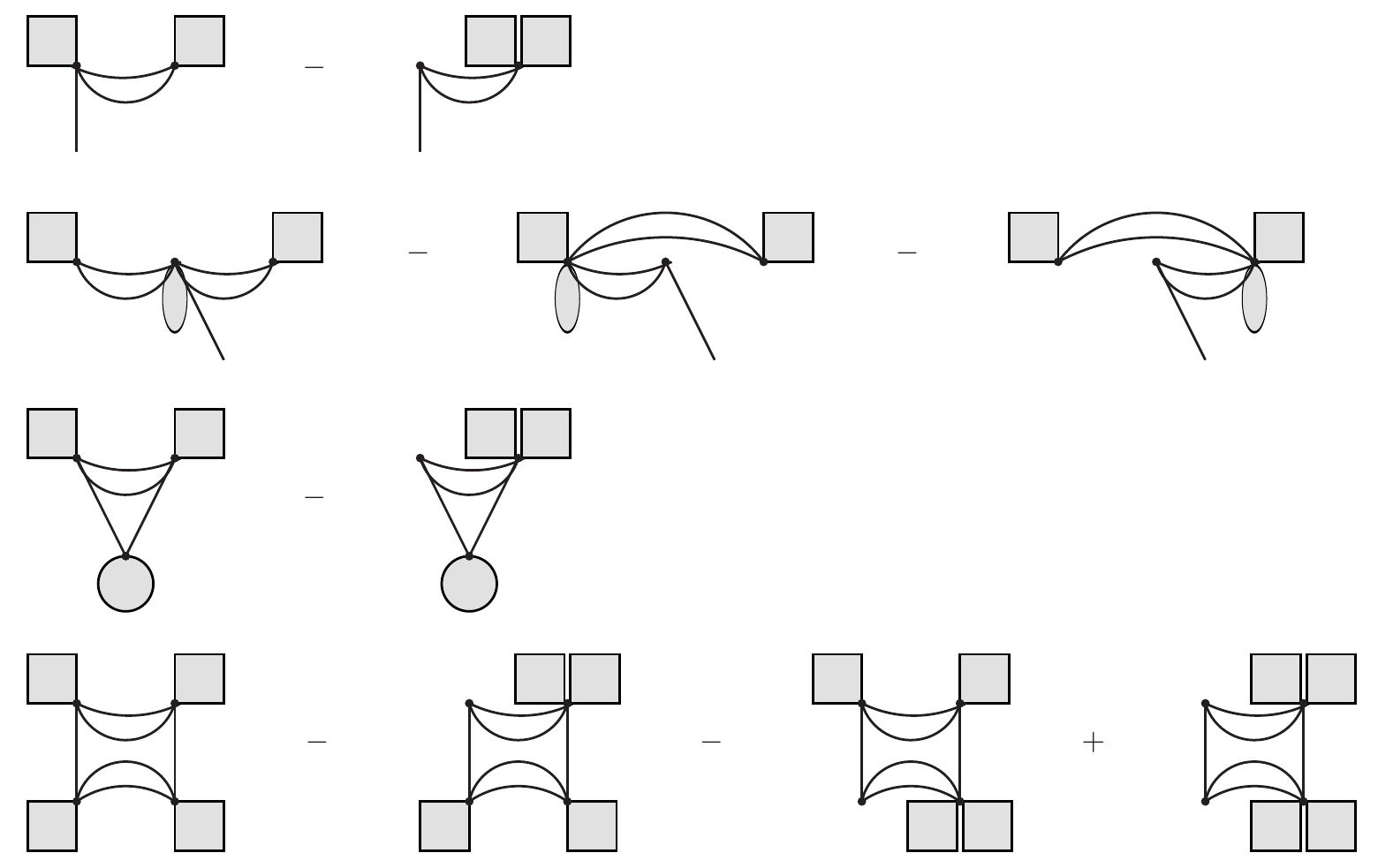}
\caption{Linear combinations of subgraphs with lifted singularities. The shaded areas connect to an ambient graph to represent a graphical function.
All but the leftmost graphs have two-chains of edges which simplify by (\ref{eqconvolute}). The central vertices of the two-chains must be internal in the full graphical function.
}
\label{fig:rerou}
\end{figure}

When using approximations (Section \ref{sectapp}) for the calculation of a graphical function, one can only handle a certain number of orders in the regulator $\epsilon$.
In this context it can be very helpful to reduce the pole order of a graphical function by subtracting subdivergences.
Subtraction of subdivergences has already been used to handle divergences for appending edges, see Section \ref{sectreg}. Here, we use a purely graphical method
which was developed by F. Brown and D. Kreimer in \cite{angels}. We change the routing of some edges in a graph $G$ in such a way that the singularity structure of the original graph
is maintained. Further, we want to generate a two-chain of edges in the subtraction graph(s), so that we can use (\ref{eqconvolute}) to eliminate the internal vertex of the
two-chain. We subtract the rerouted graph(s) from $G$ to lift the singular substructure in $G$.
It can be necessary to subtract (or add) several graphs to $G$ to fully cancel a subdivergence, see Figure \ref{fig:rerou}.
A given graphical function $f_G(z)$ can thus be expressed as a sum of graphs with a reduced subdivergence plus the subtraction graphs with a reduced vertex.
Both terms are simplified, which may allow one to compute $f_G(z)$, see Section \ref{sectex}.

A particularly accessible case occurs when one needs to know $f_G(z)$ only to order $\epsilon^{-1}$ (note that $G$ may have a high pole order due to many nested subdivergences).
In this case a complete regularization by rerouting has low degree $\epsilon^0$ and can therefore be ignored. One may thus replace $G$ by the sum of the subtraction graphs.

If a period $P_G$ (or a graphical function $f_G(z)$) needs to be calculated to order $\epsilon^0$, a complete regularization by rerouting allows one to work in integer dimensions.
Then, the period $P_G$ can be completed without generating many edges that vanish in the limit $\epsilon\to0$, see Sections \ref{sectcon} and \ref{sectapp}.

Using the renormalization Hopf algebra by A. Connes and D. Kreimer \cite{hopf,Overdiv} it is a purely combinatorial task to check if a linear combination of graphs is finite.
The reduced coproduct is given by
\begin{equation}\label{Delta}
\Delta'(G)=\sum_{g\subset G}g\otimes G/g,
\end{equation}
which extends linearly to sums of graphs. In (\ref{Delta}) the sum is over all disjoint unions $g$ of 1PI subgraphs which are logarithmically divergent by power-counting.
The graph $G/g$ is obtained from $G$ by contracting all edges of $g$. A linear combination of graphs $\Gamma$ is finite if and only if $\Delta'(\Gamma)=0$.
To show that $\Delta'(\Gamma)=0$ one can use the fact that tadpoles vanish in dimensionally regularized massless theories (one needs this e.g.\ to prove that the second line
is finite).

As an example, consider the first line $\Gamma$ in Figure \ref{fig:rerou}. For both graphs in $\Gamma$ the only 1PI subdivergent graph is the bubble. Both graphs have the same
contraction, so $\Delta'(\Gamma)=0$.

After rerouting, approximations can only be used simultaneously for all graphs in the regularized sum. The approximation is limited to the part of the ambient graph which
is not affected by the rerouting.

In {\tt HyperlogProcedures} \cite{Shlog}, subtraction by rerouting has only been implemented for simple cases. For the eight loop result of the $\phi^4$ gamma function it was
necessary to reroute some more complex graphs by hand. A general algorithm for rerouting would be desirable in the future.
Note that by three-valence, six-dimensional $\phi^3$ theory has no overlapping divergences which leads to a significantly simpler rerouting algorithm.

\section{HyperInt}\label{secthyp}
If all attempts fail to calculate a graphical function, then a last resort is parametric integration.

Graphical functions have parametric representations where the integration is over edge-variables instead of vertices \cite{par}.
A method of brute force parametric integration has been devised by F. Brown \cite{BrH1,BrH2}. The method was developed and implemented in Maple by E. Panzer
({\tt HyperInt} \cite{HyperInt}).

The method is restricted to graphical functions whose parametric representation is `linearly reducible'.
In practice, further restrictions come from severe memory and time consumption in many calculations.
The complexity of a parametric integration heavily depends on the number of edges, the order in $\epsilon$, the topology of the graph and the
weights of the edges. It proved very efficient to parametrically calculate families of graphical functions with weights $(n_e+a_e\epsilon)/\lambda$, see (\ref{nua}),
where $n_e\in\ZZ$ is fixed and $a_e\in\RR$ is a free parameter. The parameters $a_e$ can be handled by {\tt HyperInt} with moderate loss in speed and
extra memory demand. Knowledge of graphical functions with parameters can be useful for the calculation of large sets of relevant
graphical functions and Feynman periods. Note that from rather small graphical function a whole series of larger graphical functions can be constructed
with the techniques of the previous sections. Small graphical functions can thus be considered as germs which considerably enrich the set of available results.

Parametric integration in four dimensions is limited to small graphs and low orders in $\epsilon$ (typically three internal vertices with three orders in $\epsilon$,
which already needs some tweaks by the author). In six dimensions the restrictions are even more severe, so that mostly graphs with two internal
vertices were computed to low orders in $\epsilon$. This difficulty results from a higher power in the denominator of the parametric integrand and it is
a main source for the slow progress in calculating the $\phi^3$ beta function to loop order six. There exist ideas to bypass the difficulty (most prominently
by analytically pre-calculating the first integrations using general Dodgson identities \cite{BrH2}). More intense work on the level of parametric integration however has
low priority to the author of this article because it stands somewhat outside the theory of graphical functions.

\section{A worked example}\label{sectex}

\begin{figure}
\centering
\includegraphics{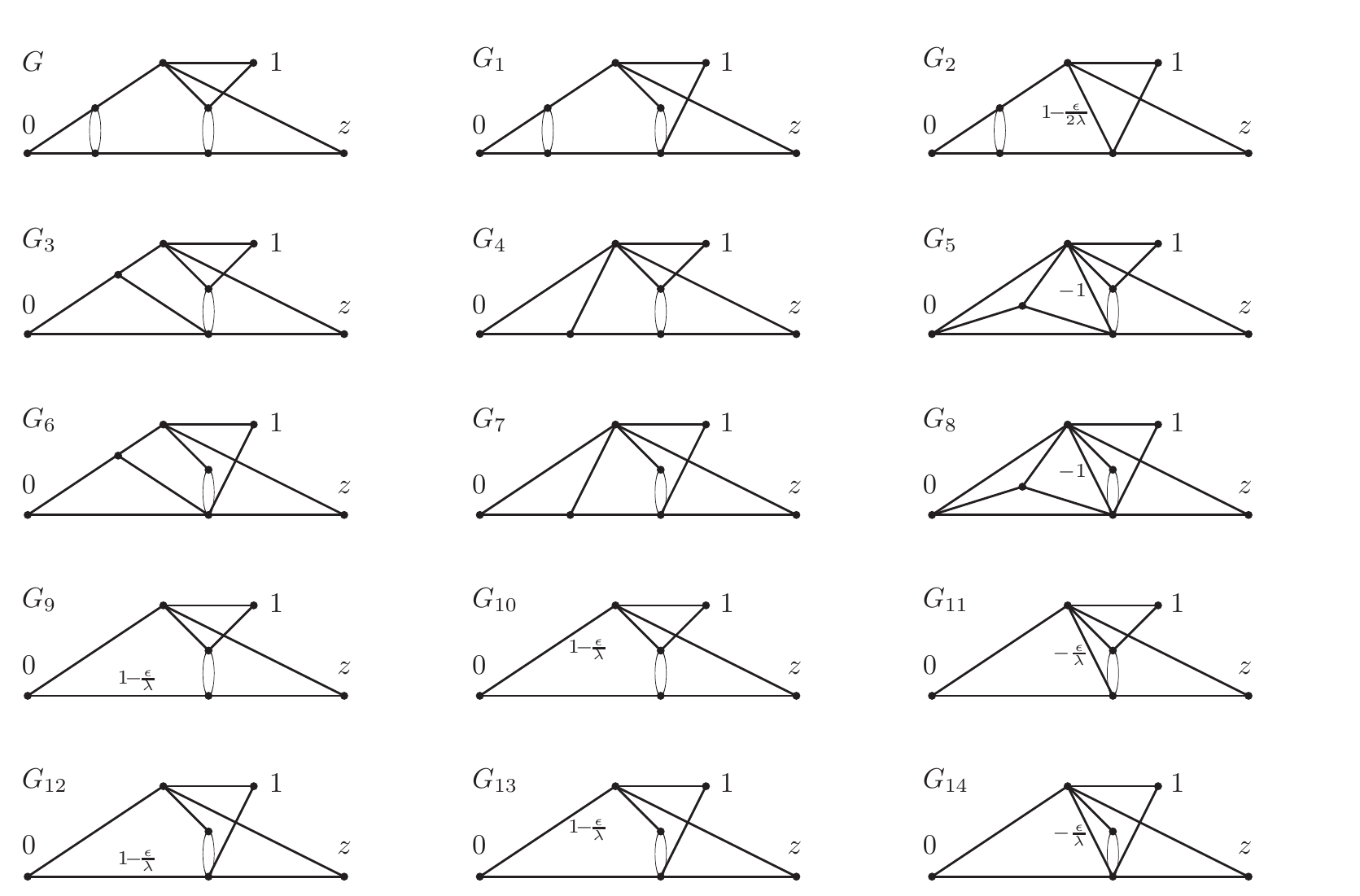}
\caption{The reduction of the graphical function $f_G(z)$ uses the graphs $G_1,\ldots,G_{14}$ in intermediate steps. Edges with no labels have weight $1$.
}
\label{fig:ex}
\end{figure}

\begin{figure}
\centering
\includegraphics{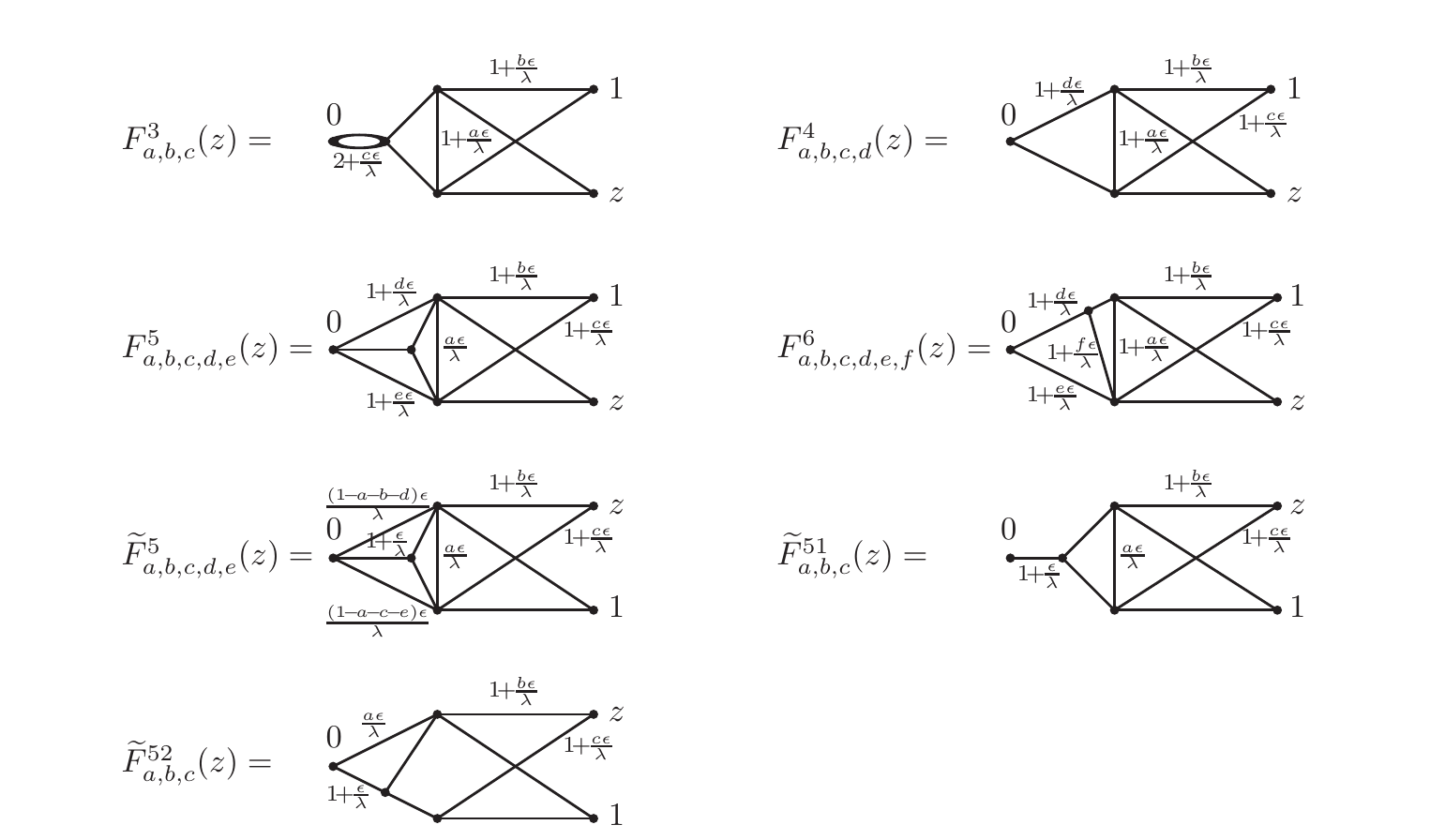}
\caption{The graphical function $f_G(z)$ in Figure \ref{fig:ex} can be expressed in terms of the graphical functions $F^3_{a,b,c}(z)$ (order $\epsilon^2$), $F^4_{a,b,c,d}(z)$
(order $\epsilon^3$), $F^5_{a,b,c,d,e}(z)$ (order $\epsilon^1$), and $F^6_{a,b,c,d,e,f}(z)$ (order $\epsilon^1$), see (\ref{fGeq}).
The graphical function $F^5_{a,b,c,d,e}(z)$ can be calculated from $\widetilde{F}^{51}_{a,b,c}(z)$ and $\widetilde{F}^{52}_{a,b,c}(z)$. Edges with no labels have weight $1$.
(The graphical functions are represented by their graphs.)
}
\label{fig:ex1}
\end{figure}

In this section we show how to combine the tools of the previous sections to calculate the graphical function of the graph $G$ on the top left of Figure \ref{fig:ex} to order $\epsilon^0$.
The graph $G$ has a triple subdivergence (Dunce's cap and a bubble) giving rise to a low degree $\epsilon^{-3}$ in $f_G(z)$.
The graph $G$ is not constructible, so we want to approximate Dunce's cap by the identities in Figure \ref{fig:app2}. We can use the second identity as it covers four orders in $\epsilon$
(from $\epsilon^{-2}$ to $\epsilon^1$). However, the second identity in Figure \ref{fig:app2} has a rather complicated first term where one loses one internal vertex but still has
to calculate four orders in $\epsilon$ (after the substitution in $G$ one needs the coefficients for $\epsilon^{-2}$ to $\epsilon^1$).
In the first equation this term is absent and the remaining terms either feature less orders in $\epsilon$ or they have lost both internal vertices. To be able to use the first identity we
regularize $G$ by subtracting the second graph $G_1$ in Figure \ref{fig:ex}. In the subtraction, the rightmost bubble is regularized (see Section \ref{sectrerou} and the first line in
Figure \ref{fig:rerou}). For the difference $f_G(z)-f_{G_1}(z)$ we only need to calculate the three coefficients of $\epsilon^{-2}$ to $\epsilon^1$. Notice that the approximation does not
use the rerouted edge. We obtain
\begin{align*}
&f_G(z)-f_{G_1}(z)=-2(f_{G_3}(z)+f_{G_4}(z)-f_{G_5}(z)-f_{G_6}(z)-f_{G_7}(z)+f_{G_8}(z))\\
&\qquad+\frac{2}{\epsilon^2(1-\epsilon)(1-\frac{3\epsilon}2)\Gamma(\lambda)^2}(f_{G_9}(z)+f_{G_{10}}(z)-f_{G_{11}}(z)-f_{G_{12}}(z)-f_{G_{13}}(z)+f_{G_{14}}(z))+\sO(\epsilon).
\end{align*}
The only singularity in the graphical functions of $G_3$, $G_4$, and $G_5$ is the bubble. They thus have pole order $-1$ and we can approximate the bubble using the first identity
in Figure \ref{fig:app1}. Using the graphical functions in Figure \ref{fig:ex1} we obtain
\begin{align*}
f_{G_3}(z)&=\frac{2F^6_{-\frac12,\frac12,-\frac12,0,0,0}(z)}{\epsilon(1-\epsilon)\Gamma(\lambda)}+\sO(\epsilon),\qquad
f_{G_4}(z)=\frac{2F^6_{-\frac12,-\frac12,\frac12,0,0,0}(z)}{\epsilon(1-\epsilon)\Gamma(\lambda)}+\sO(\epsilon),\\
f_{G_5}(z)&=\frac{2F^5_{-\frac12,\frac12,-\frac12,0,0}(z)}{\epsilon(1-\epsilon)\Gamma(\lambda)}+\sO(\epsilon).
\end{align*}
In the graphs $G_6$, $G_7$, $G_8$, $G_{12}$, $G_{13}$, and $G_{14}$ we reduce the two-chain of edges using (\ref{eqconvolute}). This gives
\begin{align*}
f_{G_6}(z)&=f_{G_7}(z)=\frac{2F^6_{-\frac12,0,0,0,0,0}(z)}{\epsilon(1-\epsilon)\Gamma(\lambda)},\qquad
f_{G_8}(z)=\frac{2F^5_{-\frac12,0,0,0,0}(z)}{\epsilon(1-\epsilon)\Gamma(\lambda)},\\
f_{G_{12}}(z)&=f_{G_{13}}(z)=\frac{2F^4_{-\frac12,0,0,-1}(z)}{\epsilon(1-\epsilon)\Gamma(\lambda)},\qquad
f_{G_{14}}(z)=\frac{2F^4_{-\frac32,0,0,0}(z)}{\epsilon(1-\epsilon)\Gamma(\lambda)}.
\end{align*}
For the graphs $G_9$ and $G_{11}$ we use the third approximation in Figure \ref{fig:app1}. In some of the resulting graphs we swap the external vertices 0 and 1 which transforms $z$ to $1-z$,
see (\ref{trafos}). This gives
\begin{align*}
f_{G_9}(z)=&-f_{G_{10}}(z)-F^3_{0,-1,0}(1-z)-\frac{\epsilon}{2(1-\epsilon)}(F^5_{0,-1,0,0,0}+F^6_{0,-1,0,0,0,0}+F^6_{0,0,-1,0,0,0})(1-z)\\
&+\,\Big(\frac2{\epsilon(1-\epsilon)\Gamma(\lambda)}+\zeta(3)\epsilon^2\Big)(F^4_{-\frac12,0,0,-1}+F^4_{0,-\frac12,0,-1}+F^4_{0,0,-\frac12,-1})(z)+\sO(\epsilon^3),\\
f_{G_{11}}(z)=&-f_{G_{11}}(z)-F^3_{-1,0,0}(1-z)-\frac{\epsilon}{2(1-\epsilon)}(F^5_{-1,0,0,0,0}+2F^6_{-1,0,0,0,0,0})(1-z)\\
&+\,\Big(\frac2{\epsilon(1-\epsilon)\Gamma(\lambda)}+\zeta(3)\epsilon^2\Big)(F^4_{-\frac32,0,0,0}+2F^4_{-1,-\frac12,0,0})(z)+\sO(\epsilon^3).
\end{align*}
In the rerouted graph $G_1$ we eliminate the two-chain using (\ref{eqconvolute}),
$$
f_{G_1}(z)=\frac{2f_{G_2}(z)}{\epsilon(1-\epsilon)\Gamma(\lambda)}.
$$
In the reduced graph $G_2$ we use the second identity in Figure \ref{fig:app2}, yielding
\begin{align*}
f_{G_2}(z)=&\,\frac3{\epsilon(1-\epsilon)\Gamma(\lambda)}F^3_{-\frac12,0,-\frac12}(z)\\
&+\frac1{4(1-\epsilon)(1-\frac{3\epsilon}{2})\Gamma(\lambda)}(2F^6_{-\frac12,0,0,-3,-\frac32,4}-F^5_{1,0,0,-1,-1})(z)\\
&-\frac1{\epsilon^2(1-\epsilon)(1-\frac{3\epsilon}{2})\Gamma(\lambda)^2}(2F^4_{1,-\frac32,-1,0}-F^4_{\frac32,-\frac32,-\frac32,0})(1-z)+\sO(\epsilon^2).
\end{align*}
Altogether, we are able to express the graphical function $f_G(z)$ in terms of the graphical functions $F^3_{a,b,c}(z)$, $F^4_{a,b,c,d}(z)$, $F^5_{a,b,c,d,e}(z)$,
and $F^6_{a,b,c,d,e,f}(z)$ in Figure \ref{fig:ex1},
\begin{align}\label{fGeq}
&f_G(z)=\frac{(8F^6_{-\frac12,0,0,0,0,0}-4F^6_{-\frac12,\frac{1}{2},-\frac{1}{2},0,0,0}-4F^6_{-\frac12,-\frac{1}{2},\frac{1}{2},0,0,0}-4F^5_{-\frac12,0,0,0,0}
 +4F^5_{-\frac12,\frac{1}{2},-\frac{1}{2},0,0})(z)}{\epsilon(1-\epsilon)\Gamma(\lambda)}\nonumber\\
&\quad+\,\frac{(F^6_{-1,0,0,0,0,0}-F^6_{0,-1,0,0,0,0}-F^6_{0,0,-1,0,0,0}+\frac{1}{2}F^5_{-1,0,0,0,0}
 -F^5_{0,-1,0,0,0})(1-z)}{\epsilon(1-\epsilon)^2(1-\frac{3\epsilon}{2})\Gamma(\lambda)^2}\nonumber\\
&\quad+\,\frac{(F^6_{-\frac{1}{2},0,0,-3,-\frac{3}{2},4}-\frac{1}{2}F^5_{1,0,0,-1,-1})(z)}{\epsilon(1-\epsilon)^2(1-\frac{3\epsilon}{2})\Gamma(\lambda)^2}
 +\frac{(F^3_{-1,0,0}-2F^3_{0,-1,0})(z)}{\epsilon^2(1-\epsilon)(1-\frac{3\epsilon}{2})\Gamma(\lambda)^2}
 +\frac{6F^3_{-\frac{1}{2},0,-\frac{1}{2}}(1-z)}{\epsilon^2(1-\epsilon)^2\Gamma(\lambda)^2}\nonumber\\
&\quad+\,\frac{4}{\epsilon^3(1-\epsilon)^2(1-\frac{3\epsilon}{2})\Gamma(\lambda)^3}\Big(\Big(1+\frac{\zeta(3)\epsilon^3}{2}\Big)
 (F^4_{0,-\frac{1}{2},0,-1}+F^4_{0,0,-\frac{1}{2},-1}-F^4_{-1,-\frac{1}{2},0,0})(z)\nonumber\\
&\qquad-\,\Big(1-\frac{\zeta(3)\epsilon^3}{2}\Big)(F^4_{-\frac{1}{2},0,0,-1}-\frac{1}{2}F^4_{-\frac{3}{2},0,0,0})(z)
 -(F^4_{1,-\frac{3}{2},-1,0}-\frac{1}{2}F^4_{\frac{3}{2},-\frac{3}{2},-\frac{3}{2},0})(1-z)\Big).
\end{align}

We need the expansions of the graphical functions $F^3_{a,b,c}(z)$, $F^4_{a,b,c,d}(z)$, and $F^6_{a,b,c,d,e,f}(z)$ to orders $\epsilon^2$, $\epsilon^3$, and $\epsilon^1$, respectively.
This can be achieved by parametric integration using (an optimized version of) {\tt HyperInt}, see \cite{HyperInt} and Section \ref{secthyp}.

The graphical function $F^5_{a,b,c,d,e}(z)$ needs to be expanded to order $\epsilon^1$. With its ten edges it is too complex to use {\tt HyperInt} directly.
We use completion (see Section \ref{sectcon}) and swap the vertices $0\leftrightarrow\infty$ and $1\leftrightarrow z$ to derive $F^5_{a,b,c,d,e}(z)$ from the graphical function
$\widetilde{F}^5_{a,b,c,d,e}(z)$ in Figure \ref{fig:ex1} (where we have deleted edges between external vertices).
In $\widetilde{F}^5_{a,b,c,d,e}(z)$ we have three edges of weights that vanish in the limit $\epsilon\to0$. Because we need $\widetilde{F}^5_{a,b,c,d,e}(z)$ only to order $\epsilon^1$ we
can expand the integrand in $\epsilon$, see Section \ref{sectexp}. This allows us to express $\widetilde{F}^5_{a,b,c,d,e}(z)$ in terms of $\widetilde{F}^{51}_{a,b,c}(z)$ and
$\widetilde{F}^{52}_{a,b,c}(z)$. Both $\widetilde{F}^{51}_{a,b,c}(z)$ and $\widetilde{F}^{52}_{a,b,c}(z)$ only have eight edges. The first two orders of their graphical functions can
easily be calculated with {\tt HyperInt}.

\section{Limitations and future developments}
At the time of writing, in the calculation of the eight loop $\phi^4$ beta function results for 19000 of 233490 Feynman periods are missing. The calculation of the six loop beta function
in $\phi^3$ theory lacks 1100 of 40788 Feynman periods. Computations are hampered by limited access to computing facilities.
Nevertheless, the author expects that a six loop result for the $\phi^3$ beta function will be available in the near future. Files of intermediate results are in \cite{Shlog}.

Difficulties in the calculation of renormalization functions are threefold:
\begin{enumerate}
\item computing time,
\item memory consumption,
\item inaccessible Feynman periods.
\end{enumerate}

Computing time is the least severe limitation. Still, a large number of Feynman periods needs to be calculated and each calculation can be quite time consuming.
In most cases the majority of computing time is used for finding a way how to do the calculation by using all implemented formulae for non-constructible Feynman
periods. One has to try all internal vertices for the external vertex $z$ and for each choice of $z$ one needs to use a large set of transformations in the attempt
to calculate the graphical function. Once a chain of transformations is found, the actual calculation typically is much faster. An exception is the case when
vast amounts of memory are used.

Memory consumption is typically rather mild (a few GB). An exception occurs if the alphabet of GSVHs uses letters that lead to numbers which are not
Euler sums. This happens quite frequently at six loops in $\phi^3$ theory. A prominent type of such numbers are iterated integrals \cite{Chen} in the alphabet
$$
X=\Big\{-1,0,\frac12,1,2\Big\}.
$$
The alphabet $X$ is a transformed (real) subset of the alphabet of sixth roots of unity $\exp(2k\pi\ii/6)$, $k=0,1,\ldots,5$. Although {\tt HyperlogProcedures} can handle
these numbers (using the `Schnetz basis'), the $\QQ$-basis of these number is quite large at high weights and transformations into the $f$-alphabet (that is used
by {\tt HyperlogProcedures} for transformation into a $\QQ$-basis \cite{MMZ,BrownDecom}) can be slow. Moreover, taking limits of such GSVHs at $z=1$ and at $z=\infty$
can be time and memory consuming.

Hitherto inaccessible Feynman periods can be calculated by using some hand-crafted reduction chains, see Section \ref{sectex}, or by generating results for graphical functions
using parametric integration, see Section \ref{secthyp}. Both approaches are cumbersome and tedious.

Systematic approaches to increase the power of the graphical functions method are the following subsections.

Moreover, the application of the graphical functions method to Gauge Theories \cite{IZ} seems possible and desirable.

\subsection{Extra identities}
The list of identities presented in this article is quite incomplete. Many identities were found in the search for a solution of a particular Feynman period.
It seems possible that many important techniques for calculating graphical functions are yet to be found.

\subsection{Recursive use of known identities}
In {\tt HyperlogProcedures} \cite{Shlog} identities are used to simplify graphical functions. In some cases, transformations into graphs of similar complexity were
accepted because the new graphs may be amenable to reductions. In the Maple implementation {\tt HyperlogProcedures} this immediately leads to very slow searches
for chains of transformations that solve a given graphical function. The search for a reduction chain is independent of the actual calculation. It can be programmed recursively
in C$++$ with much better efficiency. In the case of convergent graphical functions in six dimensions this was successfully done by M. Borinsky in \cite{phi3}. It became evident
that reduction chains often use detours over seemingly more complex graphical functions and are hence missed by {\tt HyperlogProcedures}. In particular, the IBP method (see Section
\ref{sectIBP}) which is very powerful in $\geq6$ dimensions can (by its complexity) only be handled in this approach.

The recursive calculation of graphical functions and Feynman periods in integer dimensions \cite{phi3} can be modified to compute the $\epsilon$-coefficients of regularized integrals.
It seems efficient to introduce two types of edge weights: general weights $(n_e+a_e\epsilon)/\lambda$ with $n_e\in\ZZ$, $a_e\in\RR$, as in (\ref{nua}) and special weights
$(n_e-\frac\epsilon2)/\lambda$ which, after external differentiation or integration, may become weight 1 edges that can be reduced by the appending an edge algorithm of Section \ref{sectappedge}.

Very desirable would also be a systematic implementation of rerouting, see Section \ref{sectrerou}. This method for the subtraction of subdivergences (and the reduction of the relevant
orders in $\epsilon$) has proved very efficient. Only the simplest cases are implemented in {\tt HyperlogProcedures}.

\subsection{Deep learning}
Even in a C$++$ implementation the recursive use of known identities quickly becomes too big for exhaustive results. One needs to introduce cutoffs in the size of admissible graphs
or needs to limit the set of identities which is used. The setup seems quite accessible to methods of deep learning. The power of deep learning in this context is not yet known.

\bibliographystyle{plain}

\begin{thebibliography}{99}
\bibitem{5lQCD1} {\bf P.A. Baikov, K.G. Chetyrkin and J.H. K\"uhn}, {\it Five-loop running of the QCD coupling constant}, Phys.\ Rev.\ Lett.\ 118 8, 082002 (2017).
\bibitem{5lQCD2} {\bf P.A. Baikov, K.G. Chetyrkin, J.H. K\"uhn, J. Rittinger}, {\it Vector correlator in massless QCD at order $\sO(\alpha_s^4)$ and the QED $\beta$-function at five Loop},
JHEP No.\ 7 017 (2012).
\bibitem{TMCQ} {\bf M. Borinsky}, {\it Tropical monte carlo quadrature for Feynman integrals}, to appear in Annales de l’Institut Henri Poincar\'e D, arXiv:2008.12310 [math-ph] (2020).
\bibitem{5lphi3} {\bf M. Borinsky, J.A. Gracey, M.V. Kompaniets, O. Schnetz}, {\it Five loop renormalization of $\phi^3$ theory with applications to the Lee-Yang edge
singularity and percolation theory}, Phys.\ Rev.\ D 103, 116024 (2021).
\bibitem{gfe} {\bf M. Borinsky, O. Schnetz}, {\it Graphical functions in even dimensions}, Comm.\ in Number Theory and Physics 16, No.\ 3, 515-614 (2022).
\bibitem{phi3} {\bf M. Borinsky, O. Schnetz}, {\it Recursive computation of Feynman periods}, JHEP No.\ 08, 291 (2022).
\bibitem{BrSVMP} {\bf F.C.S. Brown}, {\it Single-valued  multiple polylogarithms in one variable}, C.R. Acad.\ Sci.\ Paris, Ser.\ I 338, 527-532 (2004).
\bibitem{BrH1} {\bf F.C.S. Brown}, {\it The massless higher-loop two-point function}, Comm.\ Math.\ Phys.\ 287 925-958 (2009).
\bibitem{BrH2} {\bf F.C.S. Brown}, {\it On the periods of some Feynman integrals}, arXiv:0910.0114 [math.AG] (2009).
\bibitem{MMZ} {\bf F.C.S. Brown}, {\it Mixed Tate Motives over $\ZZ$}, Annals of Mathematics 175, No.\ 2, 949-976 (2012).
\bibitem{BrownDecom} {\bf F.C.S. Brown}, {\it On the decomposition of motivic multiple zeta values}, in Galois-Teichm\"uller theory and arithmetic geometry, vol.\ 68 of
Adv.\ Studies in Pure Math.\ (Tokyo), Math.\ Soc.\ Japan, 31-58 (2012).
\bibitem{Bcoact1} {\bf F.C.S. Brown}, {\it Feynman amplitudes, coaction principle, and cosmic Galois group}, Comm.\ in Number Theory and Physics 11, No.\ 3, 453-555 (2017).
\bibitem{Bcoact2} {\bf F.C.S. Brown}, {\it Notes on motivic periods}, Comm.\ in Number Theory and Physics 11, No.\ 3, 557-655 (2017).
\bibitem{angels} {\bf F.C.S. Brown, D. Kreimer} {\it Angles, Scales and Parametric Renormalization}, Lett.\ Math.\ Phys.\ 103, 933–1007 (2013).
\bibitem{K3} {\bf F.C.S. Brown, O. Schnetz}, {\it A K3 in $\phi^4$}, Duke Mathematical Journal, Vol. 161, No. 10, 1817-1862 (2012).
\bibitem{Cosmic} {\bf S. Caron-Huot, L.J. Dixon, F. Dulat, M. von Hippel, A.J. McLeod, G. Papathanasiou},
{\it The Cosmic Galois Group and Extended Steinmann Relations for Planar $N=4$ SYM Amplitudes}, J. High Energ.\ Phys.\ 61 (2019).
\bibitem{Duhr} {\bf F. Chavez, C. Duhr}, {\it Three-mass triangle integrals and single-valued polylogarithms}, JHEP 1211, 114-144 (2012).
\bibitem{Chen} {\bf K. Chen}, {\it Algebras of Iterated Path Integrals and Fundamental Groups}, Transactions of the American Mathematical Society, Vol.\ 156, 359-379 (1971).
\bibitem{hopf} {\bf A. Connes, D. Kreimer} {\it Hopf Algebras, Renormalization and Noncommutative Geometry}, Comm.\ Math.\ Phys.\ 199, 203–242 (1998).
\bibitem{SYM} {\bf J. Drummond, C. Duhr, P. Heslop, J. Pennington, V.A. Smirnov}, {\it Leading singularities and off-shell conformal integrals}, JHEP No.\ 8,
133-190 (2013).
\bibitem{GolzMaster} {\bf M. Golz}, {\it Graphical functions in parametric space}, Master thesis, Humboldt University Berlin, available at
{\tt http://www2.mathematik.hu-berlin.de/$\sim$kreimer/wp-content/uploads/GolzMasterThesis.pdf} (2015).
\bibitem{par} {\bf M. Golz, E. Panzer, O. Schnetz}, {\it Graphical functions in parametric space}, Lett.\ Math.\ Phys.\ 107, No.\ 6, 1177-1182 (2017).
\bibitem{5lQCD3} {\bf F. Herzog, B. Ruijl, T. Ueda, J.A.M. Vermaseren, A. Vogt}, {\it The five-loop beta function of Yang-Mills theory with fermions}, JHEP No.\ 2, 90 (2017).
\bibitem{IZ} {\bf J.C. Itzykson, J.B. Zuber}, {\it Quantum Field Theory}, Mc-Graw-Hill, (1980).
\bibitem{DY} {\bf S. Jeffries, K.A. Yeats}, {\it A degree preserving delta wye transformation with applications to 6-regular graphs and Feynman periods}, arXiv:2110.07764 [math.CO] (2021).
\bibitem{KP6loopbeta} {\bf M.V. Kompaniets, E. Panzer}, {\it Minimally subtracted six loop renormalization of $O(n)$-symmetric $\phi^4$ theory and critical exponents},
Phys.\ Rev.\ D96, No.\ 3, 036016 (2017).
\bibitem{classphi3} {\bf M.V. Kompaniets, A. Pikelner}, {\it Critical exponents from five-loop scalar theory renormalization near six-dimensions}, Phys.\ Lett.\ B 817 (2021).
\bibitem{Overdiv} {\bf D. Kreimer}, {\it On Overlapping Divergences}, Commun.\ Math.\ Phys. 204, 669-689 (1999).
\bibitem{HyperInt} {\bf E. Panzer}, {\it Algorithms for the symbolic integration of hyperlogarithms with applications to Feynman integrals}, Computer Physics Comm.\ 188, 148-166 (2015).
\bibitem{coaction} {\bf E. Panzer, O. Schnetz}, {\it The Galois coaction on $\phi^4$ periods}, Comm.\ in Number Theory and Physics 11, No.\ 3, 657-705 (2017).
\bibitem{gf} {\bf O. Schnetz}, {\it Graphical functions and single-valued multiple polylogarithms}, Comm.\ in Number Theory and Physics 8, No.\ 4, 589–675 (2014).
\bibitem{SchnetzFq} {\bf O. Schnetz}, {\it Quantum field theory over $\FF_q$}, Electron.\ J. Comb.\ 18N1:P102 (2011).
\bibitem{numfunct} {\bf O. Schnetz}, {\it Numbers and Functions in Quantum Field Theory}, Phys.\ Rev.\ D 97, 085018 (2018).
\bibitem{Shlog} {\bf O. Schnetz}, {\tt HyperlogProcedures}, Version 0.6, Maple package available on the homepage of the author, {\tt https://www.math.fau.de/person/oliver-schnetz/} (2022).
\bibitem{motg2} {\bf O. Schnetz}, {\it The Galois coaction on the electron anomalous magnetic moment}, Comm.\ in Number Theory and Physics 12, No.\ 2, 335–354 (2018).
\bibitem{Sc2} {\bf O. Schnetz}, {\it Geometries in perturbative quantum field theory}, Comm.\ in Number Theory and Physics 15, No.\ 4, 743 – 791 (2021).
\bibitem{GSVH} {\bf O. Schnetz}, {\it Generalized single-valued hyperlogarithms}, arXiv:2111.11246 [hep-th], submitted to Comm.\ in Number Theory and Physics (2021).
\bibitem{Zagierdilog} {\bf D. Zagier}, {\it The dilogarithm function in geometry and number theory}, in: Number theory and related topics. Papers presented at the Ramanujan Colloquium,
Bombay 1988, Studies in Mathematics No.\ 12, TIFR and Oxford University Press, pp.\ 231-249 (1989) and J. Math.\ Phys.\ Sci.\ No.\ 22, 131-145 (1988).
\bibitem{ZJ} {\bf J. Zinn-Justin}, {\it Quantum Field Theory and Critical Phenomena}, International Series of Monographs on Physics 113, Clarendon Press, Oxford (2002).
\end{thebibliography}
\renewcommand\refname{References}

\end{document}